\documentclass[twoside,11pt]{article}

\usepackage{feynmf}
\usepackage{epsfig}
\usepackage{umlaut}
\usepackage{tabularx}
\usepackage{amsfonts}
\usepackage{mathrsfs}
\usepackage{citesort}

\newcommand{\capskip}{\vspace*{-0.8cm}}
\newcommand{\capsize}{\small}
\newcommand{\shspace}{\hspace{0.5cm}}

\newcommand{\ba}{\begin{array}}
\newcommand{\be}{\begin{equation}}
\newcommand{\bea}{\begin{eqnarray}}
\newcommand{\bmp}[1]{\begin{minipage}{#1}}
\newcommand{\cc}[1]{{\chi_{#1}}}
\newcommand{\cb}[1]{{\bar{\chi}_{#1}}}
\newcommand{\cdef}{{:=}}
\newcommand{\cdefr}{{=:}}
\newcommand{\ce}[1]{{e^{#1}}}
\newcommand{\chib}{{\bar{\chi}}}
\newcommand{\co}{\hspace{1ex},}
\newcommand{\com}[1]{\hspace{1cm}\mbox{#1}}
\newcommand{\comm}[1]{\hspace{1cm}#1}

\newcommand{\cphi}[1]{{\phi_{#1}}}
\newcommand{\cphie}[1]{{\phi_{1,#1}}}
\newcommand{\cphiz}[1]{{\phi_{2,#1}}}
\newcommand{\cpi}{{C_\pi}}
\newcommand{\cspi}{{C_{S,\pi}}}
\newcommand{\cts}{{\mbox{CTS}}}
\newcommand{\cu}[1]{{U(#1)}}
\newcommand{\cum}[2]{{U_{#1}(#2)}}
\newcommand{\cumast}[2]{{U^\ast_{#1}(#2)}}
\newcommand{\ea}{\end{array}}
\newcommand{\ee}{\end{equation}}
\newcommand{\eea}{\end{eqnarray}}
\newcommand{\emp}{\end{minipage}}

\newcommand{\et}{\end{tabular}}
\newcommand{\hs}[1]{\hspace{#1cm}}
\newcommand{\kb}{{\bf{k}}}
\newcommand{\la}{\left(}
\newcommand{\lba}{\left[}
\newcommand{\lga}{\left\{}
\newcommand{\lla}{\left\langle\,}
\newcommand{\lts}{\mbox{LTS}}
\newcommand{\ltsS}{{{\mathscr S}_1}}
\newcommand{\ltsI}{{\mathscr I}}
\newcommand{\ltsC}{{{\mathscr C}_0}}

\newcommand{\mal}{{\bf \cdot}}

\newcommand{\mathr}{\mathbb{R}}
\newcommand{\measD}{{\cal D}}
\newcommand{\Mni}[1]{{M(#1)}}
\newcommand{\Mi}[1]{{M^{-1}(#1)}}
\newcommand{\Micc}[1]{{M^{\ast -1}(#1)}}
\newcommand{\nn}{\nonumber}
\newcommand{\om}{\omega}
\newcommand{\ord}[1]{{{\cal O}(#1)}}
\newcommand{\pb}{{\bf{p}}}
\newcommand{\Pb}{{\bf{P}}}
\newcommand{\pr}{{\prime}}
\newcommand{\psib}{{\bar{\psi}}}
\newcommand{\pu}{\hspace{1ex}.}
\newcommand{\qb}{{\bf{q}}}
\newcommand{\ra}{\right)}
\newcommand{\rba}{\right]}
\newcommand{\rga}{\right\}}
\newcommand{\rra}{\,\right\rangle}
\newcommand{\sect}{\subsection}
\newcommand{\subsect}{\subsubsection}
\newcommand{\sge}{{\sigma_1}}
\newcommand{\sgi}{{\sigma_I}}

\newcommand{\sgc}{{\sigma_C}}

\newcommand{\sgp}{{\sigma_P}}
\newcommand{\sgg}{{\sigma_G}}

\newcommand{\cslash}[1]{#1\!\!\!/}
\newcommand{\tsc}{{\mathscr C}}

\newcommand{\tsg}{{\mathscr G}}
\newcommand{\tsp}{{\mathscr P}}
\newcommand{\tsu}{{\mbox{SU(2)}}}
\newcommand{\ul}{\unitlength}
\newcommand{\xb}{{\bf{x}}}
\newcommand{\xbp}{{\bf{x}^\prime}}
\newcommand{\yb}{{\bf{y}}}
\newcommand{\ybp}{{\bf{y}^\prime}}
\newcommand{\zb}{{\bf{z}}}

\title{
\hfill \raisebox{2cm}[0cm][0cm]{\normalsize PITHA 99/13}\\[-0.8cm] 
Mass spectrum and elastic scattering in 
the massive $SU(2)_f$ Schwinger model on the 
lattice\thanks{Work supported by the Deutsche
               Forschungs\-gemein\-schaft.}
}
\author{C.~Gutsfeld$^{\rm a, b}$,
        H.~A.~Kastrup$^{\rm a}$,
        K.~Stergios$^{\rm a, b}$\\
        \small \mbox{}$^{\rm a}$ Institut f\"ur Theoretische Physik E,
                                 RWTH Aachen, 52056 Aachen, Germany\\
        \small \mbox{}$^{\rm b}$ NIC,  
                                 c/o Forschungszentrum J\"ulich, 
                                 52425 J\"ulich, Germany
       }
\date{}

\begin{document}

\maketitle

\begin{abstract}
We calculate numerically scattering phases for elastic meson--meson scattering
processes in the strongly coupled massive Schwinger--model with an
$\tsu$ flavour symmetry. These calculations are based on L\"u\-scher's
method in which finite size effects in two--particle energies are
exploited. The results from Monte--Carlo simulations with staggered
fermions for the lightest meson (``pion'') are in good agreement with
the analytical strong--coupling prediction. Furthermore, the mass spectrum
of low--lying mesonic states is investigated numerically. We find 
a surprisingly
rich spectrum in the mass region $[m_\pi,4 m_\pi]$.
\end{abstract}

\sect{Introduction}

Present experiments at high energy accelerators are mainly scattering
experiments. A task for theoretical QCD is to calculate quantities
which can be compared to the data of
such scattering experiments. 
In deep inelastic scattering processes for example,
reliable results are obtained
by perturbative calculations. But problems occur in experiments with
small momentum transfers. These are, e.g., resonant elastic scattering
processes, like the occurrence of the $\rho$--resonance and the
$\Delta$--resonance in 
$\pi$--$\pi$ and $\pi$--$p$ respectively scattering processes.
For such
low--energy phenomena perturbative calculations in the
strongly--coupled QCD region are not suitable. 
 
By contrast there exists an
appropriate method in the framework of 
lattice calculations proposed by L\"uscher in
which scattering phases of the continuum theory can be calculated for elastic
scattering processes in massive quantum field theories
\cite{Lu86b,Lu91b,Lu91a}. 
In this method one makes use of the fact that for large but finite
spatial extension the volume dependence of the energies of 
two--particle states is determined by the S--matrix for elastic scattering
processes in infinite volume. As the determination of energies in finite
volumes is possible
in Monte--Carlo simulations this method is
very useful for lattice calculations. 

There are many successful applications of this method for the
calculation of 
scattering phases in bosonic models in two and four dimensions
\cite{LW90,Ni92,GH93,GL93,GK94,GK94a,GR95}. The basic formulae for 
L\"uscher's method have also been derived in a fermionic theory
\cite{We97a}. 
The agreement of the numerical data with
the predicted scattering phases for the fermion--fermion scattering 
in the Gross--Neveu model in two dimensions \cite{GK96} 
confirms the usefulness of L\"uscher's method also in
fermionic models. 

Our aim is the application of L\"uscher's method to the determination of
elastic scattering phases in a meson--meson system of the massive
Schwinger model with an $\tsu_f$ flavour symmetry in the continuum. 
The experience
gained in this project shall support future investigations in the QCD,
e.g. of the $\rho$--resonance in the 
$\pi^+\pi^- \rightarrow \rho^0 \rightarrow \pi^+\pi^-$ 
scattering process. In this context the Schwinger model has some
useful features: 
The Schwinger model is a fermionic gauge
field theory which has many properties in common with QCD, such as
confinement, the $U(1)_A$ anomaly and a non--trivial vacuum structure. 
In particular
the massive Schwinger model with $N_f=2$ offers a complex mass
spectrum and therefore conditions (and problems) comparable to
QCD with $u$,$d$--quarks
as far as the mesonic energy spectrum is concerned.
Furthermore, the simulations for the determination of the
scattering phases are extensive, e.g. because of
the calculation of fermionic eight--point
functions. Therefore a two--dimensional model has the advantage to be
not as expensive to simulate as the
four--dimensional QCD. 

Another advantage of the massive Schwinger model with $N_f=2$ is
that there exist analytical predictions for 
strong couplings:
The massive $\tsu_f$ Schwinger model is not analytically solved. But
for strong coupling the ``pion sector'' of this model, i.e. the sector
in which the lightest meson--triplet (``pions'') occurs,
can be approximated by the sine--Gordon model. 
In the sine--Gordon model the mass spectrum and the elastic S--matrix
have been calculated \cite{DH75,Za95,ZZ79}. The primary aim of our
investigations is the comparison of the predictions for the scattering
phases from the
sine--Gordon model with the numerical results in the 
massive $\tsu_f$ Schwinger model
calculated with the procedure of L\"uscher. By this comparison we
also check ambiguities, the CDD--poles \cite{ZZ79}, 
in the analytical result for the S--matrix.  

For a successful determination of the scattering phases it is
necessary to have a good knowledge of the mass spectrum in the
massive $\tsu_f$ Schwinger model:

First, investigating the particle masses 
in the pion sector we determine the strong coupling region in which
the approximation of the pion sector by the sine--Gordon model and 
thus the prediction for the scattering phases of the pion--pion
scattering are nearly exact. Secondly, the analysis of the mass
spectrum is useful to estimate the energies of possible additional
scattering processes besides the $\pi$--$\pi$ scattering 
and of inelastic thresholds in the energy region of
the elastic $\pi$--$\pi$ scattering.   
Thirdly, we investigate systematically lattice artifacts
--- finite size and $\ord{a}$ effects --- affecting
the particle masses. Such investigations are very important 
because small systematical
deviations in the two--particle energies lead to large errors in the
scattering phases calculated from these energies.

The paper is organized as follows: 
In section \ref{schwinger} the known analytical results for the mass
spectrum and the scattering phases
are summarized. In section \ref{methods} the numerical methods, 
especially an improved method for
the determination of energies from correlation matrices, are discussed.
The relation of the symmetry groups on the lattice and in the continuum is
described in section \ref{symmetry}. 
In section \ref{massnum} the numerical results for the mass spectrum
of the massive $\tsu_f$ Schwinger model are presented. The numerical
results for the scattering phases for the 
elastic $\pi$--$\pi$ scattering are compared to the analytical
prediction in section \ref{streunum}.
In this section also the contributions of single--particle states and
of states of two non--interacting pions to four--meson correlation 
functions are discussed. The paper ends with a summary of the main results.

\sect{\sloppy Analytical predictions in the massive $\tsu_f$ Schwinger model}
\label{schwinger}
\subsect{Bosonization}
In Minkowskian space--time 
the action of the massive Schwinger model with a
$\tsu_f$ flavour symmetry is\footnote{$\hbar=c=1$,
  $g_{00}=1$, $g_{11}=-1$.}:
\bea
S & = &  \int d^2x \lga -\frac{1}{4} F_{\mu\nu} F^{\mu\nu} + \sum_{f=1,2} \psib^f ( i\cslash{\partial} - e \cslash{A}
- m_0) \psi^f\rga,  \label{schwinger-eq6}\\
& & \comm{F_{\mu\nu} = \partial_\mu A_\nu - \partial_\nu A _\mu} \,,
    \shspace \lga \gamma_\mu, \gamma_\nu \rga = 2 g_{\mu\nu} \nn \pu 
\eea
This model has many properties in common with the four--dimensional
QCD: In the massless case ($m_0=0$) 
there exists an anomalous axial $U(1)$--current
\cite{Ga96} and a massive meson (``$\eta$''--meson) 
which corresponds to the $\eta^\pr$--meson in
QCD. Also the vacuum structure of the Schwinger model
is not trivial. Analogously to the screening of
colour charges of test particles in QCD with dynamical quarks the
electric charge of test particles in the Schwinger model
is completely screened.
Furthermore, the ``quarks''
which denote the fundamental fermions in (\ref{schwinger-eq6}) are
confined in the massive Schwinger model with $N_f=2$ for $|\Theta| \ne
\pi/2$. Thus there exist no free states with a non--zero fermion number
\cite{BS79}. 

The basis for the bosonization of the Schwinger model\footnote{In the
following the term Schwinger model denotes the massive case with an
$\tsu_f$ flavour symmetry.} is
the equivalence between the sine--Gordon model and the massive
Thirring--model \cite{Co75}. 
This equivalence gives the possibility
to express the fermionic fields $\psi^1$, $\psi^2$ 
(and also the gauge fields by means of the equation of motion) in 
(\ref{schwinger-eq6}) by
two bosonic fields $\phi_+$ and $\phi_-$. 
The bosonized form of (\ref{schwinger-eq6}) which is obtained this way
was derived in ref. \cite{Co76}. 
In Hamiltonian formulation one obtains:
\bea
{\mathscr H} & = &  :\frac{1}{2}\Pi^2_+ + 
   \frac{1}{2}(\partial_1 \phi_+)^2 + \frac{\mu^2}{2}\phi^2_+:_{\mu}\:
   + : \frac{1}{2}\Pi^2_- + \frac{1}{2} (\partial_1 \phi_-)^2 :_{m_0} \nn\\
   & & - 2 c m_0^{3/2} \mu^{1/2} : \cos\la\sqrt{2\pi}\phi_+\ra :_{\mu}\: 
                                 : \cos\la\sqrt{2\pi}\phi_-\ra :_{m_0}\pu
\label{schwinger-eq9}
\eea
$:\,\,\,:_{\mu,m_0}$ denotes normal ordering with respect to
a free field with mass $\mu$ and $m_0$ respectively.
$\Pi_+$ and $\Pi_-$ are the conjugate momenta of the Bose fields
$\phi_+$ and $\phi_-$. The parameter $\mu$ is defined as $\mu \equiv
e \sqrt{2/\pi}$. 
The constant $c$ stems from
the bosonization formulae of the massive Thirring model and has the
value $c=\exp(\gamma)/(2\pi)$ \cite{GK96a} ($\gamma$ is the Euler constant).
 
We now confine ourselves
to the strong coupling limit of the model, i.e. to the case where
$m_0/e \ll 1$. 
Expanding the cosine--terms in (\ref{schwinger-eq9}) the mass scale of
the $\phi_-$-- and $\phi_+$--fields is given by 
$(m_0^{3/2}\mu^{1/2})^{1/2}$ and $\mu$ respectively. For $m_0/\mu \ll 1$
the $\phi_+$--field has a much higher mass
in (\ref{schwinger-eq9}) than the $\phi_-$--field. 
Hence for $m_0/e$ small the $\phi_+$--field has only small effects on 
the dynamics of the $\phi_-$--fields 
so that the $\phi_+$--$\phi_-$ interaction terms in 
(\ref{schwinger-eq9}) are negligible
if only the $\phi_-$--field is considered. 
With these approximations one 
finally obtains with small $m_0/e$ the following theory for the 
$\phi_-$--field \cite{Co76}:
\bea
{\mathscr H_{\phi_-}} & = & : \frac{1}{2} \Pi^2_- +
\frac{1}{2}(\partial_1\phi_-)^2 - \frac{m^{\prime 2}}{\beta_{SG}^2}
\cos(\beta_{SG}\,\phi_-) :_{m^\prime}\,, \label{schwinger-eq10}\\
& & \comm{m^\prime = 2\sqrt{2\pi} c^{2/3} \la\frac{m_0}{e}\ra^{2/3} e}
\,, \shspace \beta_{SG} = \sqrt{2\pi} \pu \nn
\eea
This Hamiltonian represents the sine--Gordon
model with a coupling parameter $\beta_{SG}$.
Hence it is possible to make use of the analytical results for the
sine--Gordon model in order to 
obtain predictions for the lightest particles in
the Schwinger model. The approximation of the Schwinger model by the
sine--Gordon model is, of course, only valid in the $\phi_-$--sector
(``pion sector'') of the Schwinger model for $m_0/e$ small.

\subsect{Mass spectrum and scattering phases} \label{predict}
The sine--Gordon model is an integrable
model. Solutions for the mass spectrum in the quantized theory have been
first derived in a semi--classical approximation \cite{DH75}. The mass
of the soliton ($A$) and antisoliton ($\bar A$) is predicted to be
\be
M_A = M_{\bar A} = \frac{8 m^\pr}{\gamma^\pr}, 
\comm{\frac{1}{\gamma^\pr} \,\cdef\, \frac{1}{\beta_{SG}^2} - \frac{1}{8\pi}}\pu
\label{schwinger-eq11}
\ee
The number of additional stable particles 
$B_n$, $n=1,2\ldots < 8\pi/\gamma^\pr$ depends on the parameter $\beta_{SG}$. 
In the case $\beta_{SG}=\sqrt{2\pi}$ there exist two stable states  
$B_1$ and $B_2$ with masses $M_1 = M_A$ and $M_2 = \sqrt{3}\,M_A$. 
Hence for $m_0/e \rightarrow 0$ there are four particles in the
Schwinger--model which are related to the $\phi_-$--sector.
The $\tsu_f$ flavour symmetry of the Schwinger
model is visible in the sine--Gordon model for
$\beta_{SG}=\sqrt{2\pi}$ as the particle spectrum consists of
a triplet ($A$,$\bar A$,$B_1$) and a singlet ($B_2$). 
The quantum numbers ($I$sospin, $P$arity, $G$--parity) of the triplet are
$I^{PG} = 1^{-+}$ and of the singlet $I^{PG} = 0^{++}$
\cite{Co76}. In analogy to QCD we call the pseudoscalar triplet
``pion'' and the $0^{++}$--particle ``$f_0$''--meson. From
(\ref{schwinger-eq10}) and (\ref{schwinger-eq11}) one obtains for the
pion mass in the Schwinger model
\bea
  m_\pi & = & 6 \sqrt{\frac{2}{\pi}} c^{2/3} \la\frac{m_0}{e}\ra^{2/3} e \nn\\
        & \simeq & 2.066 \la\frac{m_0}{e}\ra^{2/3} e \label{schwinger-eq3}
\eea
and for the $f_0$--mass
\bea
  m_{f_0} & = & \sqrt{3} \,m_\pi \pu \label{schwinger-eq4}
\eea
Exact calculations for the mass gap in the sine--Gordon model
in the full quantum field theory show 
that the semiclassical results are nearly exact
\cite{Za95,Sm97}. The difference of the exact result
\bea
  m_\pi & \simeq & 2.008 \la\frac{m_0}{e}\ra^{2/3} e
\label{schwinger-eq2}
\eea 
for the pion mass
to the semiclassical formula (\ref{schwinger-eq3}) is 
about $3\%$.

In the original derivation of the Hamiltonian (\ref{schwinger-eq10}) 
in ref. \cite{Co76} the parameter $m^\pr$ contains the additional factor
$(\cos(\Theta/2))^{2/3}$ with $\Theta \in [-\pi,\pi]$. This factor
reduces the pion mass in (\ref{schwinger-eq3}) and
(\ref{schwinger-eq2}) for $\Theta \ne 0$. 
All our Monte--Carlo
simulations yield numerical data for the pion mass which are above the
values of (\ref{schwinger-eq3}), indicating 
$\Theta \simeq 0$. Therefore we put $\Theta=0$
in all relevant formulae in this paper.

Besides the particles which have their origin in the $\phi_-$--sector
there exists a particle in the Schwinger model resulting from the 
$\phi_+$--field in (\ref{schwinger-eq9}).
It has quantum numbers $I^{PG} = 0^{--}$ (``$\eta$''--meson) \cite{Co76}.
In the massless limit $m_0 \rightarrow 0$ it becomes the massive boson of
the massless model with mass \cite{BS79}
\bea
  m = \mu \equiv e \sqrt{\frac{2}{\pi}} \pu
\label{schwinger-eq5}
\eea
In the massive theory we therefore expect for $m_0/e \ll 1$ 
a mass of the order $m_\eta \simeq \mu$. 
Other states whose mass
is above the mass of the $\pi$--,
$f_0$-- and $\eta$--meson can exist because of the $\phi_+$--$\phi_-$
interaction and the $\phi_+$ self--interactions
in (\ref{schwinger-eq9}). 
Many of such possibly unstable states have been found in
our numerical simulations (see section \ref{otherstates}). 

To obtain an analytical formula for the elastic scattering phases for the
pion--pion scattering in the strongly coupled Schwinger model one can
once again make use of its approximation by the sine--Gordon
model. The S--matrix in the sine--Gordon model has been derived
exactly \cite{ZZ79}. The calculation is based on the unitarity and
crossing--symmetry as well as on the factorization property 
of the S--matrix in the
sine--Gordon model. For $\beta_{SG}=\sqrt{2\pi}$ one obtains the same
scattering matrix element $S(\theta)$ for the three possible 
scattering processes with
soliton ($A$) and antisoliton ($\bar A$): 
$AA \longrightarrow AA$, 
$A\bar A \longrightarrow A\bar A$, 
$\bar A\bar A \longrightarrow \bar A \bar A\,$:
\bea
S(\theta) = \frac{\sinh(\theta) + i \sin{\pi/3}}{\sinh(\theta) - i \sin{\pi/3}} \pu \label{schwinger-eq1}
\eea
The difference of the rapidities $\theta_1$ and $\theta_2$ of the two
particles is denoted by $\theta$: $\theta = \theta_1 - \theta_2$. It
is related to the particle momentum $k$ by 
$\theta = 2 \,\mbox{arsinh}(k/M_A)$.  
Equation (\ref{schwinger-eq1}) is an approximation for 
the scattering matrix element for the
elastic pion--pion scattering in the strongly coupled
Schwinger model. Defining scattering phases $\delta(\theta)$ via
\bea
  S(\theta) & \equiv & \ce{2 i \delta(\theta)} \label{schwinger-eq15}
\eea  
one obtains
the connection $\delta(k/m_\pi)$ between the elastic scattering phases
and the particle momenta. The aim of our numerical investigations is
to compare the scattering phases $\delta(k/m_\pi)$ obtained
by the method of L\"uscher with
Monte--Carlo simulations with the analytical results
(\ref{schwinger-eq1}). 

The scattering matrix element (\ref{schwinger-eq1}) is
unique up to multiplicative ``CDD''--terms:
\bea
 f(\theta) = \prod_{k=1}^L 
   \frac{\sinh(\theta) + i \sin(\alpha_k)}{\sinh(\theta) - i\sin(\alpha_k)}\,,
  \comm{\alpha_k \in \mathr \pu} \label{schwinger-eq12}
\eea
Expression (\ref{schwinger-eq1}) is the ``minimal'' solution with
the lowest number of poles. Arguments that this minimal
solution for the soliton--soliton scattering in the sine--Gordon model
is exact were given in ref. \cite{ZZ79}:
There exist two states in the sine--Gordon model for
$\beta_{SG}=\sqrt{2\pi}$ with masses $M_A$ and $\sqrt{3}M_A$ which
can be interpreted as bound states of soliton and antisoliton. This
corresponds to the two poles $\theta=i\pi/3$ and $\theta=i2\pi/3$ in
the scattering matrix element 
(\ref{schwinger-eq1}) of the soliton--antisoliton
scattering. Furthermore, the minimal solution is in agreement with the
results for the S--matrix in the semiclassical limit
$\beta_{SG} \rightarrow 0$. Our numerical results discussed in section
\ref{streunum-results} 
also support the scattering phases of the minimal solution.

Because of the factorization property of the S--matrix the scattering
matrix elements of scattering processes involving the states $B_n$ can be
calculated by using the scattering matrix element of the soliton--soliton
scattering.
The states $B_n$ can be interpreted as bound states
of soliton and antisoliton. 
Hence arbitrary scattering processes
can be understood as a sequence of scattering processes between
soliton and antisoliton. Calculating for $\beta_{SG} = \sqrt{2\pi}$
the scattering matrix elements for the processes 
$B_1 A \rightarrow B_1 A$, 
$B_1 \bar A \rightarrow B_1 \bar A$ and
$B_1 B_1 \rightarrow B_1 B_1$
one obtains in each case the expression (\ref{schwinger-eq1}). Hence
all scattering processes within the triplet ($A$, $\bar A$, $B_1$)
have the same scattering matrix element. 
That means that the elastic scattering phases
for the $\pi$--$\pi$--scattering in the Schwinger model for 
$m_0/e \rightarrow 0$
are identical for all isospin channels.

\sect{Numerical methods} \label{methods}

Our starting point for Monte--Carlo simulations of the
Schwinger model is its Euclidean lattice action
with Kogut--Susskind fermions: 
\bea
S & = & S_W + \sum_{x,y} \cb{x}  \Mni{x,y} \cc{y} \,, \\
S_W & = & \beta \sum_x \lga 1 - \mbox{Re} \,U_p(x) \rga \,, \nn\\ 
   & & \shspace \shspace 
       U_p(x) \,\cdef \,U^\ast_2(x) U^\ast_1(x+\hat 2) U_2(x+\hat 1)
    U_1(x)\,, \nn\\
\Mni{x,y} & = & \frac{a}{2} \sum_\mu  \eta_\mu(x) 
        \lga  \cu{x,x+\hat\mu} \delta_{x+\hat\mu,y} -  \cu{x,x-\hat\mu} \delta_{x-\hat\mu,y} \rga \nn\\
        & & + m_0 a^2 \delta_{x,y} \,, 
        \shspace \shspace \eta_1(x) = 1\,,\,\, \eta_2(x) = (-1)^{x_1} \pu  
        \label{methods-eq2}
\eea 
The sum over the space--time coordinates $x=(x_1,x_2)$ runs for the
space coordinate  $x_1 \equiv \xb$ from $a$ to $L a$ and for the time
coordinate $x_2 \equiv t$ from $a$ to $T a$. 
We use the Wilson action with the link variables
$U_\mu(x)$ as a compact
formulation of the gauge fields: $U_\mu(x) \equiv U(x+\hat \mu,x) \in U(1)$. 
The Kogut--Susskind fermions $\chib,\chi$ on the lattice have one
flavour so that
a two--flavour Schwinger model in the continuum is simulated.
In the simulations we use periodic boundary conditions for the 
link variables and periodic boundary 
conditions in space and antiperiodic in time for the fermionic fields.

In the naive continuum limit the parameter $\beta$ has to be chosen 
$\beta=1/(e^2 a^2)$ to obtain the continuum action of the
Schwinger model. Therefore for finite $a$ we define 
a coupling $e$ on the lattice
by $e \,\cdef \,\beta^{-1/2}$. Here and in the following the lattice
constant $a$ is set to one. Dimensional
and dimensionless parameters and fields are denoted by the same 
symbols provided there is no confusion.

\subsect{HMC and topological ergodicity} \label{hmc}

\begin{figure}[tb]
\begin{minipage}[b]{0.5\textwidth}
\hspace*{-0.5cm}\epsfig{file=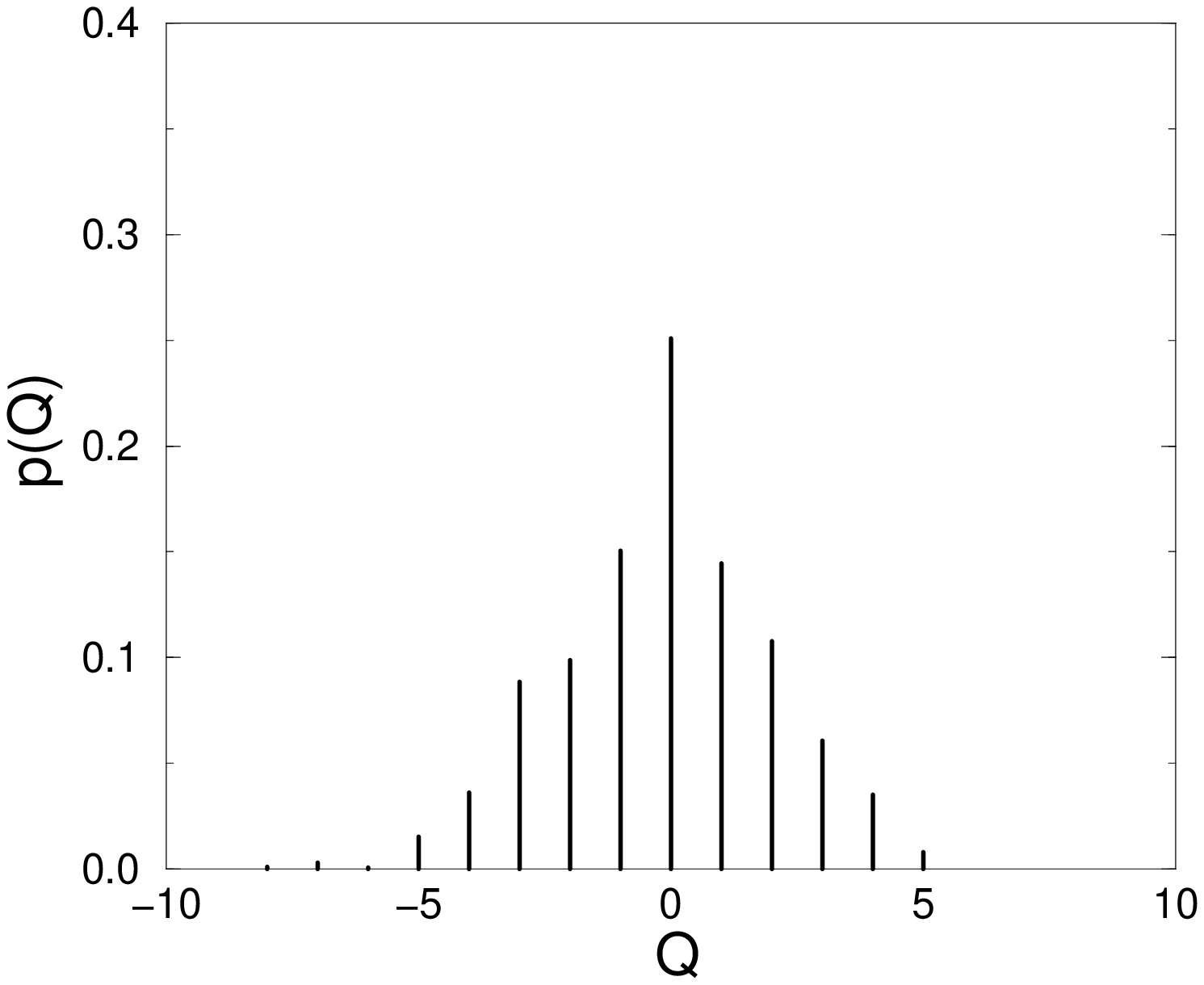,width=1.1\textwidth}
\emp
\hspace*{-0.5cm}
\begin{minipage}[b]{0.5\textwidth}
\hspace{-0.5cm}\epsfig{file=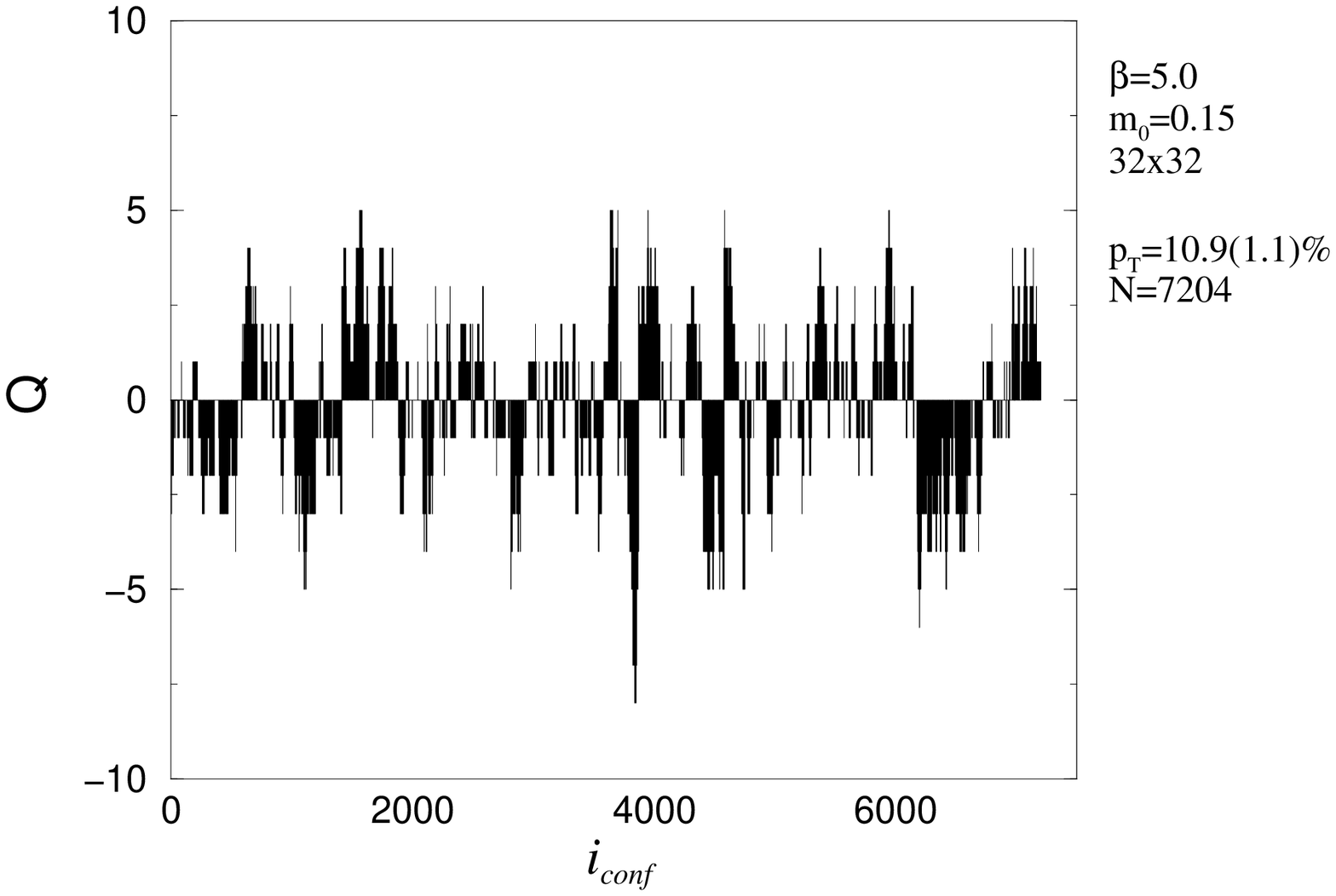,width=1.1\textwidth}
\emp\\
\capskip
\caption{\capsize Probability distribution 
and time series of the topological
charge $Q$ for a typical simulation with $N=7204$ configurations for 
($\beta=5$, $m_0=0.15$).} 
\label{methods-fig1}
\end{figure}

We use the Hybrid Monte Carlo (HMC) method with pseudofermions and the
conjugate gradient algorithm for the inversion of the fermion matrix.
The HMC provides
an efficient algorithm for the calculation of Euclidean path integrals 
in the Schwinger model. 
For $\beta \le 5$ it appears to be ergodic with respect
to the transition between different topological sectors
(``topological ergodicity''). 
Topological sectors are
characterized by the topological charge $Q$ of the gauge field
configurations. We use the 
geometrical definition of refs. \cite{Lu82,Pa85}:
\bea
   Q = \frac{1}{2\pi} \sum_x \theta_p(x) \pu
\eea 
$\theta_p \in (-\pi,\pi]$ is the plaquette angle which is defined by
$U_p(x) = \exp(i \theta_p)$. The tunneling probability $p_T$, i.e. the
probability for a change of the topological sector in the
HMC simulations, is about $10\%$ for
$\beta = 5$ and for ten trajectories between the measurements, 
reaching an
acceptance probability of about 60-70\%.
The probability distribution of the
topological charge $p(Q)$ for this coupling is plotted in fig. 
\ref{methods-fig1}. 
Its shape can be understood
phenomenologically from
investigations in the pure $U(1)$ theory \cite{GL97}. 
The mean action (and also the minimum action) in each topological sector
is proportional to $Q^2$ on large lattices. Assuming a Boltzmann
distribution one obtains a Gaussian like shape for $p(Q)$ as in
fig. \ref{methods-fig1}.       
The difference of the mean (and minimum) value of the action for adjacent
topological sectors increases for increasing $\beta$. Therefore the
width of $p(Q)$ becomes smaller for larger $\beta$ values so that for 
$\beta \ge 10$ almost only the sector $Q=0$ contributes. 

In the
quasi microcanonical calculation of the trajectories 
in the Hybrid Monte Carlo the action serves as a potential.  
Hence one is confronted with the problem that for high $\beta$--values 
the tunneling between adjacent topological sectors is suppressed.
In fact the tunneling probability $p_T$ for $\beta \ge 10$
is far below $0.1\%$.
In that case the HMC algorithm is not suitable to perform simulations
having topological ergodicity, which is also known from QCD.
This is, however, only a practical
problem for the calculation of ensembles of moderate size which
becomes less relevant on very large ensembles.  

Another problem
follows from the fact that the correlation functions have different 
values in different topological sectors: In contrast to 
investigations
for $\beta=1$ \cite{GK98} it turns out that for high values of $\beta$ the
pion mass differs significantly for different topological
sectors \cite{St99a}. 
Also the exponential decay of the correlation function
$C(t)$ for the pion is modified for $Q \ne 0$ 
(see ref. \cite{EB97} for Wilson--fermions). For
large $t$ and $|Q|$ the values of $C(t)$ are substantially 
smaller than it is expected for a pure
exponential decay. Thus it is only possible for small values of $t$ 
and $|Q|$
to determine the pion mass by fits with exponential functions.
But one is interested in high $\beta$--values as the
continuum limit $a \rightarrow 0$ of the Schwinger model requires 
$\beta \rightarrow \infty$. A possible solution to the
problems is the introduction of improved HMC algorithms which force the
tunneling between different topological sectors \cite{Di95,FH98}. 

We have chosen an alternative way by calculating expectation
values for $\beta \ge 10$ only for $Q=0$. We started 
the Monte--Carlo simulations with a suitable gauge configuration with
$Q=0$. Because of the small $p_T$ for $\beta \ge 10$ the generation of
ensembles of $\ord{50000}$ was possible without tunneling into
sectors $Q\ne 0$. 
This procedure is justified because the probability
distribution $p(Q)$ of the
topological charge for $\beta \ge 10$ is quite narrow. Hence the
expectation value of an operator $O$ which can be decomposed on the
lattice as
\bea
  \lla O \rra = \sum_{\nu = -\infty}^\infty \lla O \,\delta_{Q[U],\nu} \rra
\label{methods-eq11}
\eea
is dominated by the $Q=0$ sector:
\bea
  \lla O \rra & \simeq & \lla O\,\delta_{Q[U],0} \rra \pu
  \label{methods-eq4}
\eea
This approximation is valid as long as the absolute size 
of the expectation values for $Q \ne 0$ in (\ref{methods-eq11}) 
is much less than for $Q=0$. A possible reason for large expectation
values for $Q \ne 0$ 
could be the influence of approximate zero modes of the fermion matrix
$M$ on fermionic correlation functions like
\bea
 \lla \cb{x} \cc{x} \cb{y} \cc{y} \rra & = &
 \lla \Mi{x,x} \Mi{y,y} - \Mi{x,y} \Mi{y,x} \rra_U \nn \co \\
 & & \lla \ldots \rra_U \equiv \frac{1}{Z} \int \measD U \det M \ldots
 \ce{-S_W} \co \nn \\
 & & \hspace{1.7cm} Z = \int \measD U \det M \,\ce{-S_W} \nn
\eea
with the consequence that the elements of the inverse fermion matrix
become large. But as the fermion matrix for $m_0 = 0$ is antihermitian for
staggered fermions it has no zero modes for $m_0 \ne 0$. Therefore we
do not expect large contributions for $Q \ne 0$ for the bare masses
$m_0 \ne 0$ we
used in our simulations. (The same problem for the massless
Schwinger model with staggered fermions was discussed in
ref. \cite{Di93}.)
With the approximation (\ref{methods-eq4})
the quantity $\lla O \rra$ can be calculated in two ways:
Either the 
expectation value of $O\,\delta_{Q[U],0}$ for all possible gauge field
configurations is determined or the calculation of $O$ is performed with the
probability density
\bea
  P_{Q=0} = \frac{1}{Z} \delta_{Q[U],0} \, \ce{-S} \pu
  \label{methods-eq5}
\eea
In the latter case the acceptance probability for
configurations with $Q[U] \ne 0$ is zero. 
Thus only the ergodicity of the algorithm within the
topological sector $Q=0$ is needed. In real simulations the explicit
rejection of a configuration according to the $\delta$--function in
(\ref{methods-eq5}) is
not necessary for $\beta \ge 10$. As it was mentioned above the
tunneling probability $p_T$ for high $\beta$ is so low
that no tunneling occurs for ensembles of moderate size. 

A known disadvantage of HMC simulations for high $\beta$--values is
that due to weakly fluctuating gauge fields the autocorrelation of
gauge and fermionic operators increases. Therefore for $\beta=10$ we
have calculated ten HMC trajectories between the measurements to obtain
uncorrelated values for the interesting operators. 

\subsect{Enlargement of correlation matrix}

The main task of the analysis of the numerical data is the extraction
of energies from connected correlation matrices 
$C_{ij}(t) = \langle O_i(t) O_j(0)\rangle_c$, $i,j = 1\ldots r$.
A standard method for this is the calculation of generalized
eigenvalues of the matrix $C(t) \equiv (C_{ij}(t))$:
\bea 
  C(t) w^{(l)}(t) =  \lambda_l(t) C(t_0) w^{(l)}(t) \co & & l=1 \ldots
  r  \co \label{methods-eq1}\\ 
  & & t_0 \le t  \co \nn
\eea 
The generalized eigenvalues $\lambda_l(t)$ have 
the following form for large
times ($t\rightarrow \infty$) \cite{LW90}:
\bea 
  \lambda_l(t) = (\sigma_l)^{t-t_0} \ce{-E_l (t-t_0)} +
  \ord{\ce{-\Delta E_l (t-t_0)}}\,,\comm{\sigma_l = \pm 1} \pu
\label{methods-eq10}
\eea 
$\Delta E_l$ is the smallest difference between the
energy $E_l$ and other energies in the spectrum:
$\Delta E_l = \min_{j \ne l} \left| E_j - E_l \right|$ . The prefactor
$(\sigma_l)^{t-t_0}$ is due to the usage of staggered fermions.
If $\sigma_l = +1,-1$ we call the eigenvalues and energies
non--alternating and alternating respectively.
 
One possibility to relate the numerically found eigenvalues 
$\lambda(t)$ for different times $t$ to a fixed $l$ is to consider 
the absolute size of the eigenvalues.
This may lead to problems if the energies in the 
spectrum and hence the eigenvalues are close to each other.
A procedure which turns out to be often
more successful is the assignment
by means of the generalized eigenvectors
$w^{(l)}$: 
The generalized eigenvectors $w^{(l)}(t^\prime)$ 
of a timeslice $t^\pr$ (normally $t^\pr = t_0+1$)
are reference vectors
for the assignment of the eigenvalues. A generalized eigenvalue
$\lambda_{\pi(l)}(t)$ on the time slice $t$ is assigned to the eigenvalue
$\lambda_l(t^\pr)$ if $\pi \in S^r$ is a permutation which fulfills:
\bea
  \sum_{l=1}^r \left| w^{(l)}(t^\pr) \cdot w^{(\pi(l))}(t) \right| \hs{0.5}\mbox{maximal} \pu
\eea
This means that the eigenvectors of the time slice $t$ should be chosen as
parallel as possible to the reference eigenvectors of the time slice
$t^\pr$. 

\begin{figure}[tb]
\vspace*{-1.5cm}
\hspace*{1cm}
\epsfig{file=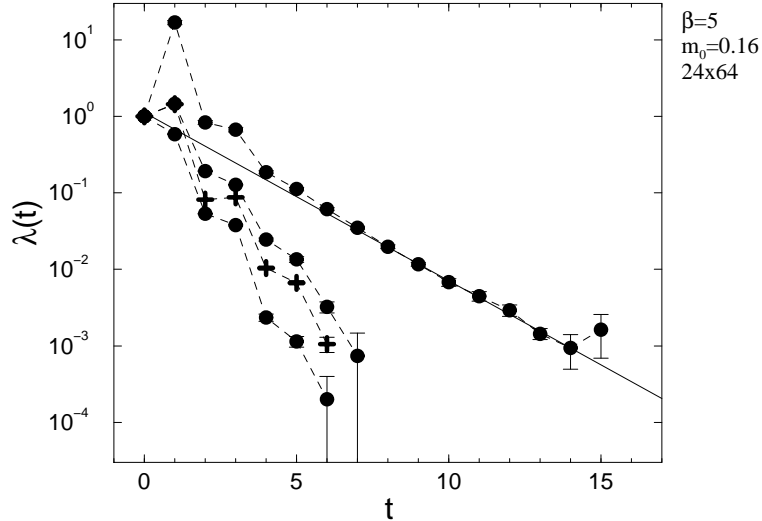,height=0.5\textheight}\\
\capskip
\caption{\capsize Logarithmic plot of the generalized eigenvalues of a
  four--meson correlation matrix $C_{ij}(t)$, $i,j=1\ldots
  4$. Positive values are represented by circles and negative values
  by crosses.}
\label{methods-fig4}
\end{figure}

In some cases even the above methods are not 
sufficient to extract energies from a correlation matrix.
In figure \ref{methods-fig4} the generalized eigenvalues for a typical
$4 \times 4$ four--meson correlation matrix are plotted. 
It is apparent
that the different eigenvalues are close to each other and that they
have time oscillating corrections which are due to alternating energies. 
This
could lead to a wrong assignment of the eigenvalues. Apart from the
highest eigenvalue it is also not possible to perform a fit with a
single exponential function. Analysis also for other parameter values
shows that there are only slight improvements if the number of
operators $O_i(t)$ is increased.

In such cases a substantial progress can be made by {\em
enlarging} the correlation matrices. In this (to our knowledge)
new method a correlation matrix is enlarged to
its double size to reduce the contributions of higher energies:

Let $C(t)$ be a (not necessarily symmetric) 
$r \times r$ correlation matrix with elements $C_{ij}(t)$ of the form:
\bea
        C_{ij}(t)  =  \sum_{k=1}^{2r} v_i^{\pr(k)} v_j^{(k)}
        (\sigma_k)^t \ce{-E_k t} \label{methods-eq9} \pu
\eea
A $(2r) \times (2r)$--matrix $E(t)$ is now constructed by 
\bea
        E_{2i-1+\tilde{i}, 2j-1+\tilde{j}}(t) & \cdef &
        C_{ij}(t+\tilde{i}+\tilde{j}), \label{methods-eq6} \\
        & & i,j = 1 \ldots r, \nn \\
        & & \tilde{i},\tilde{j} = 0,1 \nn \pu
\eea
E.g., a $4 \times 4$ matrix $E(t)$ which is made up
of the elements of a $2\times 2$ matrix $C(t)$ has the following
form:
\bea E(t) & = & \la 
\ba{ll|ll} C_{11}(t) & C_{11}(t+1) & C_{12}(t) & C_{12}(t+1) \\
           C_{11}(t+1) & C_{11}(t+2) & C_{12}(t+1) & C_{12}(t+2) \\\hline   
           C_{21}(t) & C_{21}(t+1) & C_{22}(t) & C_{22}(t+1) \\
           C_{21}(t+1) & C_{21}(t+2) & C_{22}(t+1) & C_{22}(t+2) 
\ea
\ra \pu \nn \eea
Defining vectors ($e^{\pr(k)},e^{(k)}$) via
\bea
        E_{i^\pr j^\pr}(t)  \cdefr   \sum_{k=1}^{2r}
        e_{i^\pr}^{\pr(k)} e_{j^\pr}^{(k)} (\sigma_k)^t \ce{-E_k t}
        & & \forall t,\comm{i^\pr,j^\pr = 1 \ldots 2r} \co
        \label{methods-eq7}
\eea
one obtains for these vectors with definition (\ref{methods-eq6}):\\
\bmp{.4\textwidth}
\bea 
        e_{2i-1}^{\pr(k)} & = & v_i^{\pr(k)}, \nn \\
        e_{2i-1}^{(k)} & = & v_i^{(k)}, \nn 
\eea
\emp
\bmp{.4\textwidth}
\bea
        e_{2i}^{\pr(k)} & = & \sigma_k \ce{-E_k} v_i^{\pr(k)}, \nn \\
        e_{2i}^{(k)}    & = & \sigma_k \ce{-E_k} v_i^{(k)} \pu \nn
\eea
\emp
\bmp{.2\textwidth} \bea \nn\\ \label{methods-eq8} \eea \emp\\[0.5cm]
The form (\ref{methods-eq7}) of the elements of the matrix $E(t)$ is
suitable to
determine the energies $E_k$, $k=1\ldots 2r$ by the calculation of
the generalized eigenvectors of $E(t)$. For these calculations
it is required that the vectors $e^{\pr (k)}$,
$e^{(k)}$ are linearly independent. This has to be checked from case to
case during the numerical analysis.

\begin{figure}[tb]
\vspace*{-1.5cm}
\hspace*{1cm}
\epsfig{file=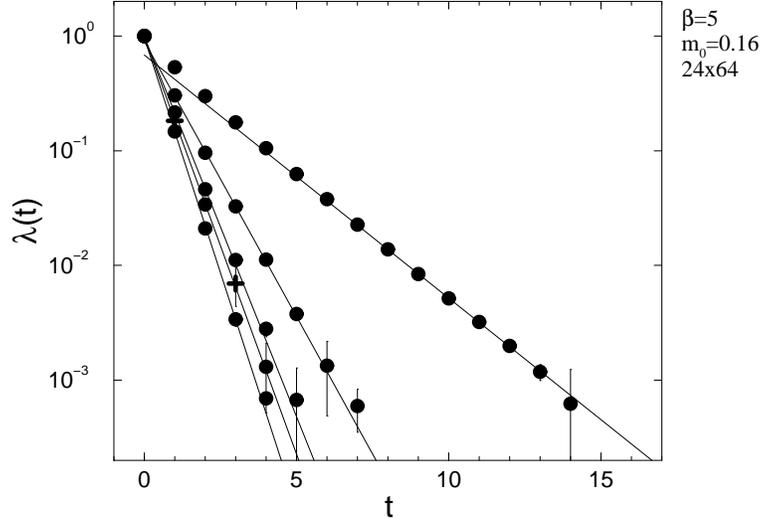,height=0.5\textheight}\\
\capskip
\caption{\capsize Generalized eigenvalues of an enlarged $4\times 4$
  correlation matrix. Plotted are those eigenvalues which correspond
  to the ones in fig. \ref{methods-fig4} as well as an additional alternating
  eigenvalue.} 
\label{methods-fig5}
\end{figure}

The main advantage of the enlargement of the correlation matrices is
that according to (\ref{methods-eq8}) the amplitudes 
$e_j^{(k)}$, $e_j^{\pr(k)}$, $\,j=2,4,\ldots,2r$ are weighted
with the factor $\exp(-E_k)$. 
This results in a suppression 
of higher energies in the spectrum by a stronger exponential factor
in (\ref{methods-eq7}).
Hence one obtains a smaller distortion
of lower eigenvalues through
higher energies and a better separation of the generalized eigenvalues. 
Furthermore, on all time slices a better differentiation
between alternating ($\sigma_k=-1$) and non--alternating
($\sigma_k=+1$) eigenvalues is possible
as the eigenvectors $e^{(k)}$ change qualitatively for $\sigma_k=-1$
because of the multiplication of each second element with $\sigma_k$.
The success of this method is obvious in fig. \ref{methods-fig5} which is the
analogue
of fig. \ref{methods-fig4} for an enlarged correlation matrix.
An improvement of the determination of the eigenvalues and energies 
is obvious. A further effect is the doubling of the number of 
determinable energies for a
correlation matrix with fixed size. 

In real applications the time extension $T$ of the lattices is finite
and a replacement of exponential functions by hyperbolic function in 
(\ref{methods-eq9}) is necessary. The adjustment of the enlargement
method
in this case is done by a modification of the equations
(\ref{methods-eq6}):
\bea
E_{2i-1, 2j-1}(t) & = & C_{ij}(t)\,, \nn \\
E_{2i, 2j-1}(t) & = & C_{ij}(t-1) + C_{ij}(t+1)\,, \nn \\
E_{2i-1, 2j}(t) & = & E_{2i, 2j-1}(t)\,, \nn \\
E_{2i, 2j}(t) & = & C_{ij}(t-2) + C_{ij}(t+2) - 2 C_{ij}(t)
\eea 
and (\ref{methods-eq8}):\\  
\bmp{.4\textwidth}
\bea
        e_{2i-1}^{\pr(k)} & = & v_i^{\pr(k)}, \nn \\
        e_{2i-1}^{(k)}    & = & v_i^{(k)}, \nn 
\eea
\emp
\bmp{.4\textwidth}
\bea
        e_{2i}^{\pr(k)}   & = & \sigma_k \cosh(-E_k) v_i^{\pr(k)}, \nn\\
        e_{2i}^{(k)}      & = & \sigma_k \cosh(-E_k) v_i^{(k)} \pu \nn 
\eea
\emp
\bmp{.2\textwidth} \bea \nn\\  \eea \emp\\[0.5cm]
Five matrices $C(t-2),C(t-1),\ldots,C(t+2)$ are needed per
constructed matrix $E(t)$. Therefore the number of time slices on which 
the matrix $E(t)$ can be evaluated is reduced by four. 
Concerning the matrix
enlargement one also has to be aware of the fact that in comparison
to the matrix $C(t)$ larger statistical errors for the generalized
eigenvalues of $E(t)$ have to be taken into account. 
The reason is that the signal/noise--ratio of a general correlation matrix
$C(t)$ decreases for larger times $t$. As the correlation matrix $C(t+2)$
with its statistical error contributes to the matrix $E(t)$
the signal to noise ratio of $E(t)$ is in many cases
worse than that of the matrix
$C(t)$. This is true as far as the matrices $C(t)$ are statistically 
uncorrelated for different time slices $t$. If they are statistically
correlated the opposite effect might occur. 
In that case the calculation of the generalized eigenvalues
can result in a cancellation of the statistical fluctuations of the
matrices $C(t)$.
This might lead to an 
underestimation of the statistical error. Therefore a careful check
of the statistical errors is necessary when applying the enlargement
method.

\sect{Symmetry transformations} \label{symmetry}

In all simulations we used time slice operators in which the
dependence on the space coordinate drops out by summation.
Therefore we are interested in those symmetry transformations of the action
which leave the time $t$ invariant. We use in particular the 
``bosonic'' representation $D$, which is defined by its action on
gauge--invariant, mesonic operators --- composed of fermion,
antifermion and gauge field ---  with total momentum $\Pb =0$. In this
representation transformations that multiply fermion and
antifermion with phase factors $\exp(-i \phi)$ and $\exp(i
\phi)$ respectively are trivial. Such symmetries which are not trivial in this
representation form the continuum time slice group $\cts$ and
the lattice time slice group $\lts$ respectively. 
In the following a short description
of the groups in two dimensions and the connection between lattice and
continuum irreducible representations is given. The procedure to
connect the lattice and continuum groups is presented in more details
in refs. \cite{KS87,GS84,GS85,G86}.  

The relevant symmetries of the continuum action of the
Schwinger model are
\begin{itemize}
\item parity $\tsp$,
\item G--parity $\tsg$
\begin{minipage}[t]{.5\textwidth} 
\vspace{-0.93cm}
  \bea 
     & \cdef & \ce{i\frac{\pi}{2}\sigma_2} \tsc \nn\\
     & = & i \sigma_2 \tsc, \com{$\tsc$ charge conjugation,}\nn
  \eea
\emp\\
\item $\tsu_f$ flavour transformation.
\end{itemize}
The G--parity is defined as charge conjugation with subsequent
rotation in isospin space. 
The above symmetry transformations for the continuum action
commute. They are the generators of the free
group $\cts=\lla \tsp\, \tsg \, \tsu_f \rra$. 
Following the notation of ref. \cite{GK95} we characterize an irreducible
bosonic representation (also called symmetry sector)
of the $\cts$ by $\bar{\Delta}^{\sgp\sgg}_{\bar{D}}$. 
The quantum numbers of the discrete transformations $\tsp$ and $\tsg$ 
are denoted
by $\sgp,\sgg=\pm 1$ and the irreducible representation of the
$\tsu_f$ group by $\bar{D}$. 

For the Schwinger model on the lattice with staggered fermions
the following symmetry transformations exist:
\begin{itemize}
  \item shift in space--direction $\ltsS$, 
  \item spatial inversion $\ltsI$,
  \item charge conjugation $\ltsC$.
\end{itemize}
The exact action of these symmetry
transformations
on the fermions and gauge fields is described in
ref. \cite{KS87}. 
According to the continuum case we denote an irreducible bosonic 
representation of the group $\lts = \lla \ltsS\, \ltsI \, \ltsC \rra$ by 
\bea
  \Delta^{\sge \sigma_I\sigma_C} \equiv (\sigma_1,\sigma_I,\sigma_C) \pu
  \label{sym-eq3}
\eea
In the following mainly the short notation on the right--hand
side of equation (\ref{sym-eq3}) is used. 

The ansatz for the embedding of the symmetry group of the
$\lts$ in the $\cts$ are the equations
\bea
  \psi^f_\mu (2 x) & = & 2^{-3/2} \sum_H \gamma^{(H)}_{\mu f}
  U(2 x,2 x + \hat H) \,\cc{2 x + \hat H}\,, \label{sym-eq1} \\
  \psib^f_\mu (2 x) & = & 2^{-3/2} \sum_H \cb{2 x + \hat H} \, U(2 x+\hat H,2
  x) \gamma^{(H)^\ast}_{ \mu f} \co \label{sym-eq2} \\
  & & H = 0,1,2,12\co \,\, \hat{H} = 0,\hat{1},\hat{2}, \hat{1} +
      \hat{2}\co \,\, \gamma_H = 1,\gamma_1,\gamma_2,\gamma_1\gamma_2 \co\nn
\eea
which connect the fermionic fields on the lattice $\chib,\chi$ with
those of the continuum $\psib,\psi$ \cite{KM83}.
By applying the symmetry transformations of the $\lts$ on the fields
$\psi$ in terms of the lattice field $\chi$ one obtains transformation
properties of the continuum fields. From this one obtains the
connection between the $\cts$ and the $\lts$.

For the determination of the mass spectrum we use mesonic operators,
i.e. tensors of rank two
\be
  \psi^f_{f^\pr} \equiv \psib^f \psi^{f^\pr}
  \label{sym-eq5}
\ee
and for the scattering phases tensors of rank four 
\be
  \psi^{ij}_{kl} = \psib^i \psi^k \psib^j \psi^l \pu 
  \label{sym-eq6}
\ee
It is known that the irreducible representations of the $\cts$
restricted on the subgroup $\lts$ are reducible and decompose
into a direct sum of irreducible representations of the $\lts$.
For the tensors (\ref{sym-eq5}) and (\ref{sym-eq6}) 
the connection between the irreducible
representations is listed in table \ref{einb} which is the corrected
version of table $1$ in ref. \cite{GK98}. 
\begin{table}[tb]
\setlength{\tabcolsep}{0.2pc}

\begin{tabular}{|ccc||ccc|ccc|ccc|ccc|ccc|ccc|}\hline 
  \multicolumn{3}{|c||}{LTS} & \multicolumn{18}{c|}{CTS} \\
  \hline
  \multicolumn{3}{|c||}{} & \multicolumn{6}{c|}{} & 
    \multicolumn{12}{c|}{}\\
  \multicolumn{3}{|c||}{$\Delta^{\sigma_1\sigma_I\sigma_C}$} & 
    \multicolumn{6}{c|}{$\bar{\Delta}_{\bar{D}}^{\sigma_P\sigma_G}$
    (rank 2)} & 
    \multicolumn{12}{c|}{$\bar{\Delta}_{\bar{D}}^{\sigma_P\sigma_G}$
    (rank 4)}\\
  \multicolumn{3}{|c||}{} & \multicolumn{6}{c|}{}  & 
    \multicolumn{12}{c|}{}\\
  $\sigma_1$ & $\sigma_I$ & $\sigma_C$ & $\bar{D}$ & $\sigma_P$ & 
    $\sigma_G$ & $\bar{D}$ & $\sigma_P$ & $\sigma_G$ & $\bar{D}$ & 
    $\sigma_P$ & $\sigma_G$ & $\bar{D}$ & $\sigma_P$ & $\sigma_G$ & 
    $\bar{D}$ & $\sigma_P$ & $\sigma_G$ & $\bar{D}$ & $\sigma_P$ & 
    $\sigma_G$ \\ 
  \hline  
  --&--&--&{ 1 }&+ &--&{ 1 }&--&--&
    { 2 }&--&--&{ 2 }&+ &--&
    { 1 }&--&--&{ 1 }&+ &-- \\
  --&--&+ &{ 1 }&+ &+ &{ 1 }&--&+ &
    { 2 }&--&+ &{ 2 }&+ &+ &
    { 1 }&--&+ &{ 1 }&+ &+ \\
  --&+ &--&{ 1 }&--&--&{ 1 }&+ &-- &
    { 2 }&--&--&{ 2 }&+ &--&
    { 1 }&--&--&{ 1 }&+ &-- \\
  --&+ &+ &{ 1 }&--&+ &{ 1 }&+ &+ &
    { 2 }&--&+ &{ 2 }&+ &+ &
    { 1 }&--&+ &{ 1 }&+ &+ \\
  + &--&--&{ 1 }&+ &--&{ 0 }&--&--&
    { 2 }&--&--&{ 2 }&+ &--&
    { 1 }&+ &--&{  0 }&--&-- \\
  + &--&+ &{ 1 }&+ &+ &{ 0 }&--&+ &
    { 2 }&--&+ &{ 2 }&+ &+ &
    { 1 }&+ &+ &{  0 }&--&+  \\
  + &+ &--&{ 1 }&--&--&{ 0 }&+ &-- &
    { 2 }&--&--&{ 2 }&+ &--&
    { 1 }&--&--&{  0 }&+ &--  \\
  + &+ &+ &{ 1 }&--&+ &{ 0 }&+ &+ &
    { 2 }&--&+ &{ 2 }&+ &+ &
    { 1 }&--&+ &{  0 }&+ &+  \\
  \hline
\end{tabular}

\caption{\capsize Connection between the irreducible representations 
         of CTS and LTS.}
\label{einb}
\end{table}
Each lattice sector is connected with two and four respectively different
continuum sectors.
 
A better assignment of an energy determined in a definite lattice
sector to a sector in the continuum is obtained by using 
the results of the transfer matrix formalism for staggered fermions 
(see ref. \cite{AB93} and the references therein).
With these results the factors $\pm (\pm 1)^t$ in the 
eigenvalues of the correlation matrices $C(t)$ can be related to
discrete quantum numbers of the $\lts$ and $\cts$.
Hence for tensors of rank two a unique assignment
of the energies obtained for a definite lattice sector to a state in a
continuum sector is possible.

In contrast to the singlet states there exist three lattice sectors
for tensors of rank two 
in which a triplet with given continuum quantum numbers can be
investigated. Therefore it is much more easier to find a suitable
operator for the investigation of the triplet states, e.g. the pion. 
According to 
\bea
  1^{-+} \otimes 1^{-+} & = & 2^{++} \oplus \, 1^{++} \oplus \, 0^{++}
  \nn
\eea
a two--pion state exists in three different isospin channels. 
This leads to four different lattice sectors in which a priori 
investigations for the $\pi$--$\pi$ scattering can be performed. 
As it will be shown in the discussion of the numerically determined
scattering phases in section \ref{streunum} there exists a posteriori 
only one
lattice sector which is suitable for the determination of the elastic
$\pi$--$\pi$ scattering phases.

\sect{Mass spectrum} \label{massnum}

In the context of this paper the mass spectrum of the Schwinger model
is important for two reasons.
First, the lightest particles offer
the opportunity to check the analytical predictions 
(in the strong coupling limit) for the mass spectrum 
in the Schwinger model with numerical methods.
Secondly, no exact
calculation for the complete mass spectrum of the massive
Schwinger model exists up to now. But for the
interpretation of the numerically determined scattering phases 
it is necessary to have
information about the lightest particles in the spectrum. Particles with
masses $m>2m_\pi$ might occur as resonances in the elastic
$\pi$--$\pi$ scattering whereas other mesons with $m<2m_\pi$ might cause
inelastic processes at energies below the threshold $E = 4 m_{\pi}$.

\subsect{Correlation functions} \label{masscorr}
The simplest time slice operator which can be used for the calculation
of the pion mass is
\bea
 O(t) & = & \frac{1}{L} \sum_\xb \cb{x} \cc{x} (-1)^x \,,
 \shspace (-1)^x \equiv (-1)^{\xb+t} \pu
\eea
The easiest way to calculate the correlation function $C(t) = \langle
 O(t) O(0) \rangle$ for the investigation of a triplet 
 is to consider only the connected part in the correlation function:
\bea
  \cpi(t) \equiv \frac{1}{L^2} \sum_{\xb,\yb} \lla |\Mi{x,y}|^2 \rra_U \pu
\eea  
It is expected that for exact flavour symmetry in the continuum limit,
the $C_\pi$--function yields the correct pion energies \cite{Ke86}.

For the investigation of other mesonic states we used a
general two--fermion operator:
\bea
O(t) & = & \sum_{\xb} \cphi{\xb} \cb{\xb,t} \cc{\xb +
  \kappa_1,t+\tau_1} \cu{\xb,t;\xb + \kappa_1,t+\tau_1}, \label{massnum-gl4} \\
 & & \comm{\cphi{\xb} = \phi_1^\xb,\,\, \phi_1 = \pm 1,} 
     \comm{\kappa_1,\tau_1 = 0,1} \nn \pu
\eea
Charge conjugation eigenstates are constructed in the usual
way by using a linear combination of (\ref{massnum-gl4}) and 
its charge conjugate. The possible
values for $\kappa_1$,$\tau_1$ and $\phi_1$ are restricted by the requirement
that the operator should also be an eigenstate of the symmetry
transformations $\ltsS$ and
$\ltsI$. To reduce the complexity of the correlation functions and
hence the computational effort we did not construct eigenstates of
the inversion $\ltsI$ 
for all quantum number combinations. 
Nevertheless by considering the results of the
transfer matrix formalism it is in all cases possible to determine the
continuum quantum numbers of the measured energies
(see section \ref{symmetry}).

Because of the complexity of the mass spectrum we calculated 
correlation matrices:
\bea
  C_{ij} (t) = \lla O_i(t) O_j(0) \rra \pu
\label{massnum-eq12}
\eea
The operators $O_i(t),\,i=1\ldots r$,
differ only by different values of the smearing parameters. 
We use smeared sources with the smearing
function $S_{\xb_0}(\xb)$ calculated by a Jacobi iteration \cite{Al93}:
\bea
   S_{\xb_0}^{(0)}(\xb)  & \cdef & \delta_{\xb,\xb_0} \,,\nn\\ 
   S_{\xb_0}^{(n)}(\xb)  & \cdef & \frac{1}{1+2\alpha} \left(
     S_{\xb_0}^{(n-1)}(\xb) + \alpha \left\{ S_{\xb_0}^{(n-1)}(\xb+2) +
       S_{\xb_0}^{(n-1)}(\xb-2) \right\} \right) \pu \nn\\
     \label{massnum-eq11}
\eea
Parallel transporters have to be inserted into (\ref{massnum-eq11}) 
so that the function $S_{\xb_0}(\xb)$ obtains the correct
gauge transformation properties. After $N$ iterations the normalized smearing
function $S_{\xb_0}(\xb)$ with approximate Gaussian shape is defined by  
\bea 
   S_{\xb_0}(\xb) & \cdef & S_{\xb_0}^{(N)}(\xb) / \left\| S_{\xb_0}^{(N)} \right\| \,,
        \shspace \left\|S_{\xb_0}\right\| \cdef 
        \left[ \sum_\xb S_{\xb_0}^\dag(\xb) S_{\xb_0}(\xb)
        \right]^{1/2} \pu \nn
\eea
The calculation of this function only on even/odd sites is necessary
for KS--fermions \cite{Fr97a}. The use of smeared sources is most
successful for the calculation of the pion mass. With $N=20$ which is
chosen throughout and $\alpha=0.1$ it is possible to suppress strongly
other states in the $C_\pi$--function. The radius $r$ of the Gaussian
curve $S_{\xb_0}(\xb)$ defined by
\bea
        r^2  & \cdef & \sum_\xb \xb^2 \, S^\dag_{\xb_0}(\xb) \, S_{\xb_0}(\xb)
\eea
in that case is $r \simeq 2.5\,$.

The different values of $\alpha$ for the operators in
(\ref{massnum-eq12}) are chosen below $\alpha=0.1$.
The one--particle energies are extracted from 
the generalized eigenvalues of the
correlation matrix $C_{ij}(t)$. Because of the
rich mass spectrum it was in most cases necessary to enlarge the
correlation matrix.

\subsect{Finite size effects and dispersion relation}
Besides the knowledge of the mass spectrum 
a good control over the lattice artifacts is
necessary as far as the elastic scattering phases are
concerned. The 
finite size effects which are used to determine the
scattering phases have to be separated from unwanted 
exponentially suppressed finite size effects (polarization effects). 
Furthermore, the correct dispersion
relation has to be chosen to calculate the particle momenta of the
scattering particles (see section \ref{streunum-luescher}) and hence 
the scattering phases. 

For massive scalar bosonic quantum field theories L\"uscher derived
a formula for the difference of a particle mass in finite ($m_L$)
and infinite ($m_\infty$) volume \cite{Lu84a,Lu86a}. For the lightest
particle in the spectrum one obtains in two dimensions:
\bea
\Delta m & = & m_L - m_\infty  \nn \\
         & \simeq  & - F(0) \frac{1}{m_\infty} \frac{1}{4} \frac{1}{\sqrt{2\pi
           L_{eff}}} \ce{-L_{eff}}\co \comm{L_{eff} \,\cdef \,L \,m_\infty} \pu
\label{massnum-eq5}
\eea 
The scattering amplitude $F(\nu)$ depends on the variable $\nu = (\om(\pb)
\om(\qb) - \pb \qb)/m_\infty$ with $\pb$ and $\qb$ being the momenta
of the incoming particles and $\om(\pb)$ the relativistic mass of the
particles:
\bea
  \om(\pb) = \sqrt{m_\infty^2 + \pb^2} \pu
\eea   
Equation (\ref{massnum-eq5}) is valid provided that no one--particle
exchange scattering processes
occur in the quantum field theory. This is fulfilled in the
sine--Gordon model for $\beta_{SG} = \sqrt{2\pi}$. From the definition
of $F(\nu)$ in ref. \cite{Lu84a} one obtains with $\nu = m_\infty
\cosh(\theta)$ a connection
between $F(\nu)$ and the scattering matrix element $S(\theta)$ in
(\ref{schwinger-eq1}). For $\nu = 0$ one obtains $F(0)
\propto - m_\infty^2$. Inserting this in (\ref{massnum-eq5}) one obtains
\bea
  \Delta m & = & + A \frac{\sqrt{m_\infty}}{\sqrt{L}} \ce{-L_{eff}}
\label{massnum-eq4}
\eea
with a numerical prefactor $A$. Considering the connection between the
sine--Gordon model and the pion sector of the Schwinger model
we use the formula (\ref{massnum-eq4}) as a possible approximation 
to describe the mass shift of the pion in the Schwinger model
due to finite size corrections.
Because of additional states in the spectrum the formula 
(\ref{massnum-eq4}) which we use for fits is not as theoretically 
justified as L\"uscher's formula (\ref{massnum-eq5}).

\begin{figure}[p]
\begin{minipage}[b]{0.5\textwidth}
\hspace*{-0.5cm}\epsfig{file=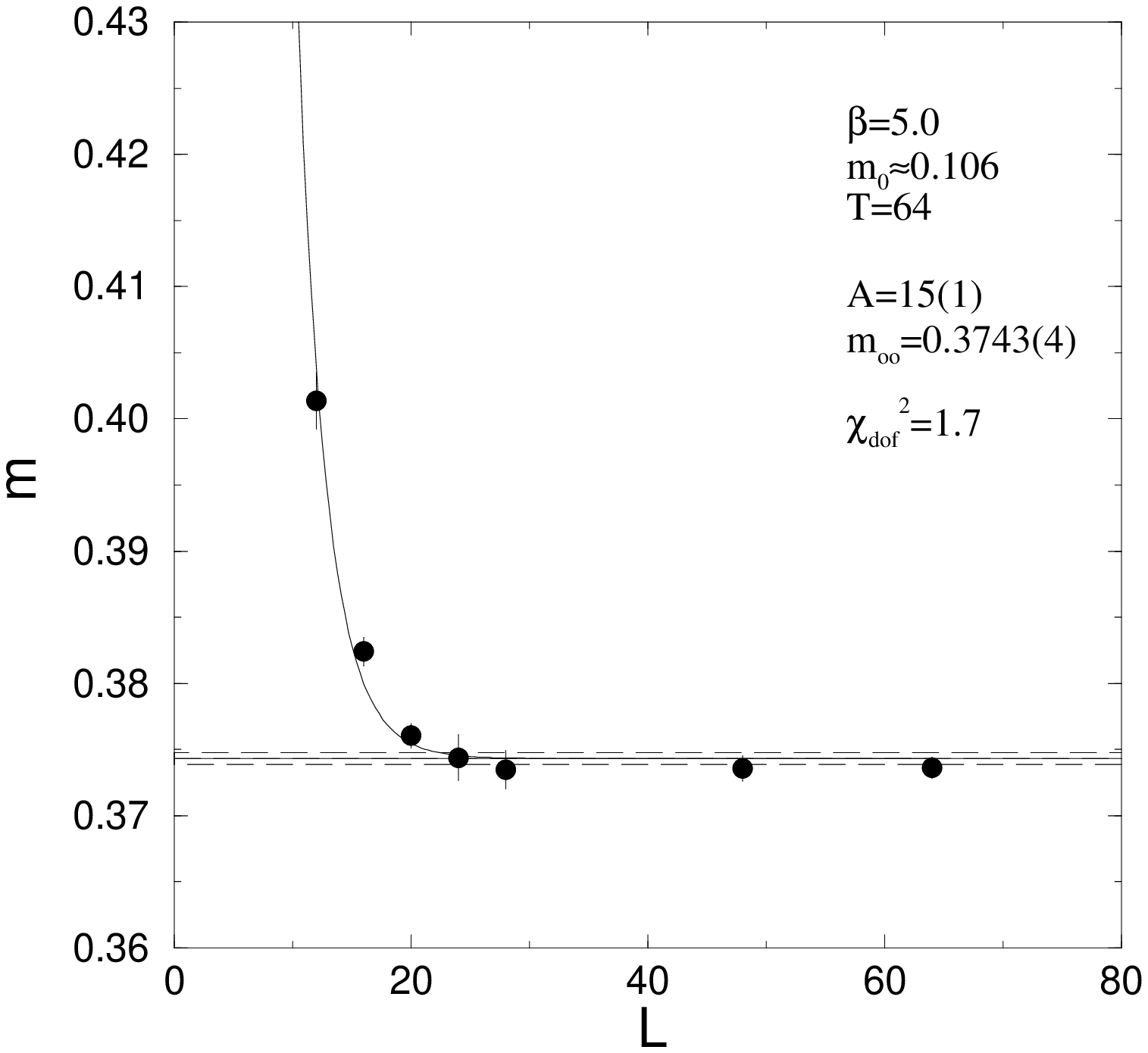,width=1.2\textwidth}
\emp
\begin{minipage}[b]{0.5\textwidth}
\hspace*{-0.6cm}\epsfig{file=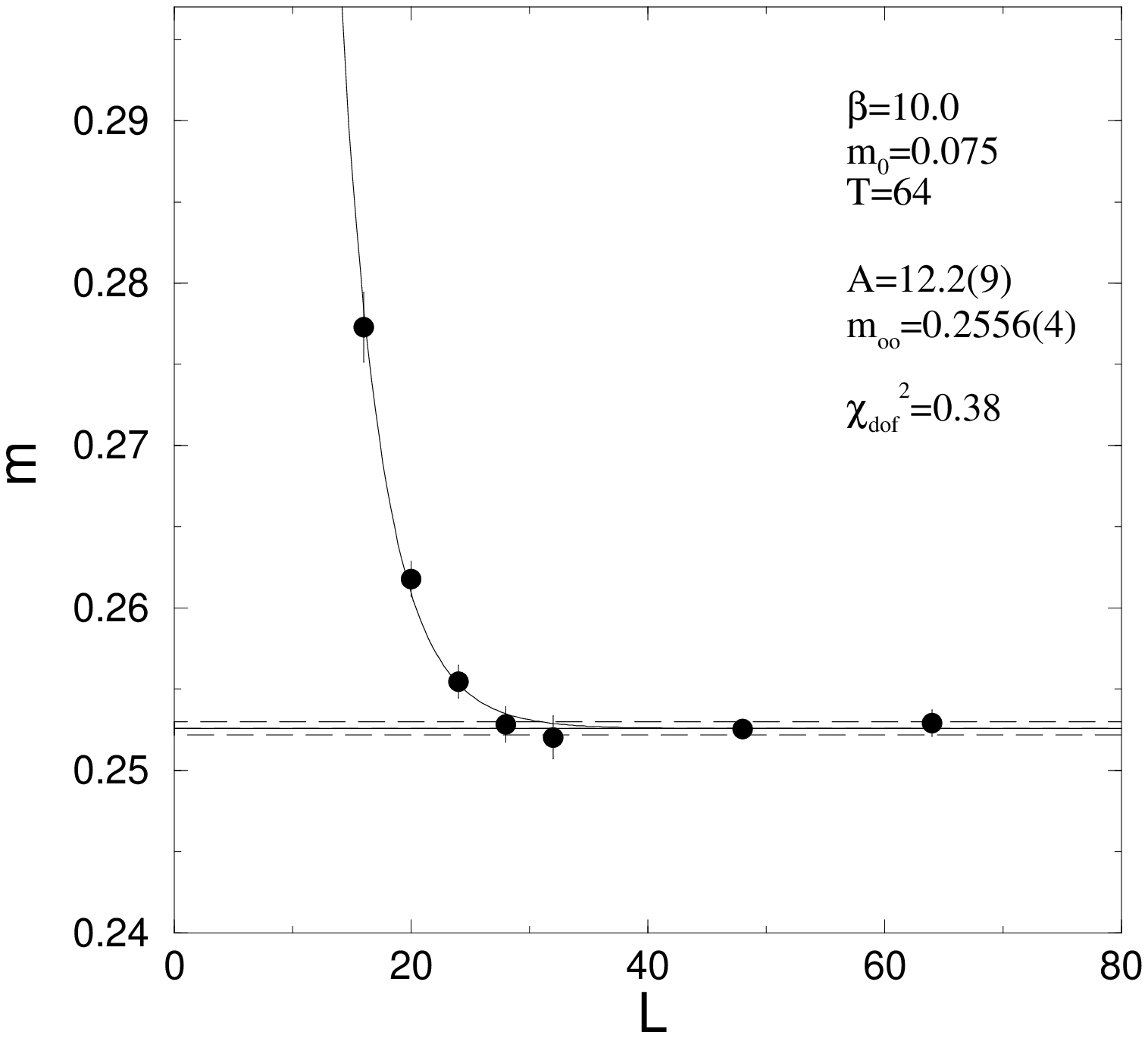,width=1.2\textwidth}
\emp\\
\capskip
\caption{\capsize The pion mass as a function of the spatial lattice
extension for two different couplings $\beta$ and fixed $m_0
\sqrt{\beta} = 0.075 \sqrt{10}$. The mass in the
infinite volume $m_\infty$ results from a fit with the function
(\ref{massnum-eq4}) to the numerical data.}
\label{massnum-fig2}
\end{figure}

The numerical results 
for the finite size effects on the pion mass
are shown in fig. \ref{massnum-fig2} for $\beta=5$ and $\beta=10$.
The value of the ratio $m_0/e = m_0 \sqrt{\beta}$ 
is fixed so that 
corrections to $F(0)$ with respect to $m_0/e$ 
and therefore to the amplitude $A$ 
are the same in both cases. A fit to the data in terms
of the ansatz (\ref{massnum-eq4}) 
shows a good agreement between
this formula and the numerical data with similar amplitudes for
both couplings (see fig. \ref{massnum-fig2}). It turns out
that strong finite size effects exist for $L_{eff} \le 6$. 
For $L_{eff} \ge 8$ the mass shifts are far below $1\%$. 
In fig. \ref{massnum-fig2}
they are comparable to the statistical
errors for $L_{eff}=8$.
Therefore for all Monte--Carlo simulations we have chosen the
spatial extension so that the effective length is $L_{eff} \ge 8$.

\begin{figure}[p]
\begin{minipage}[b]{0.5\textwidth}
\hspace*{-0.5cm}\epsfig{file=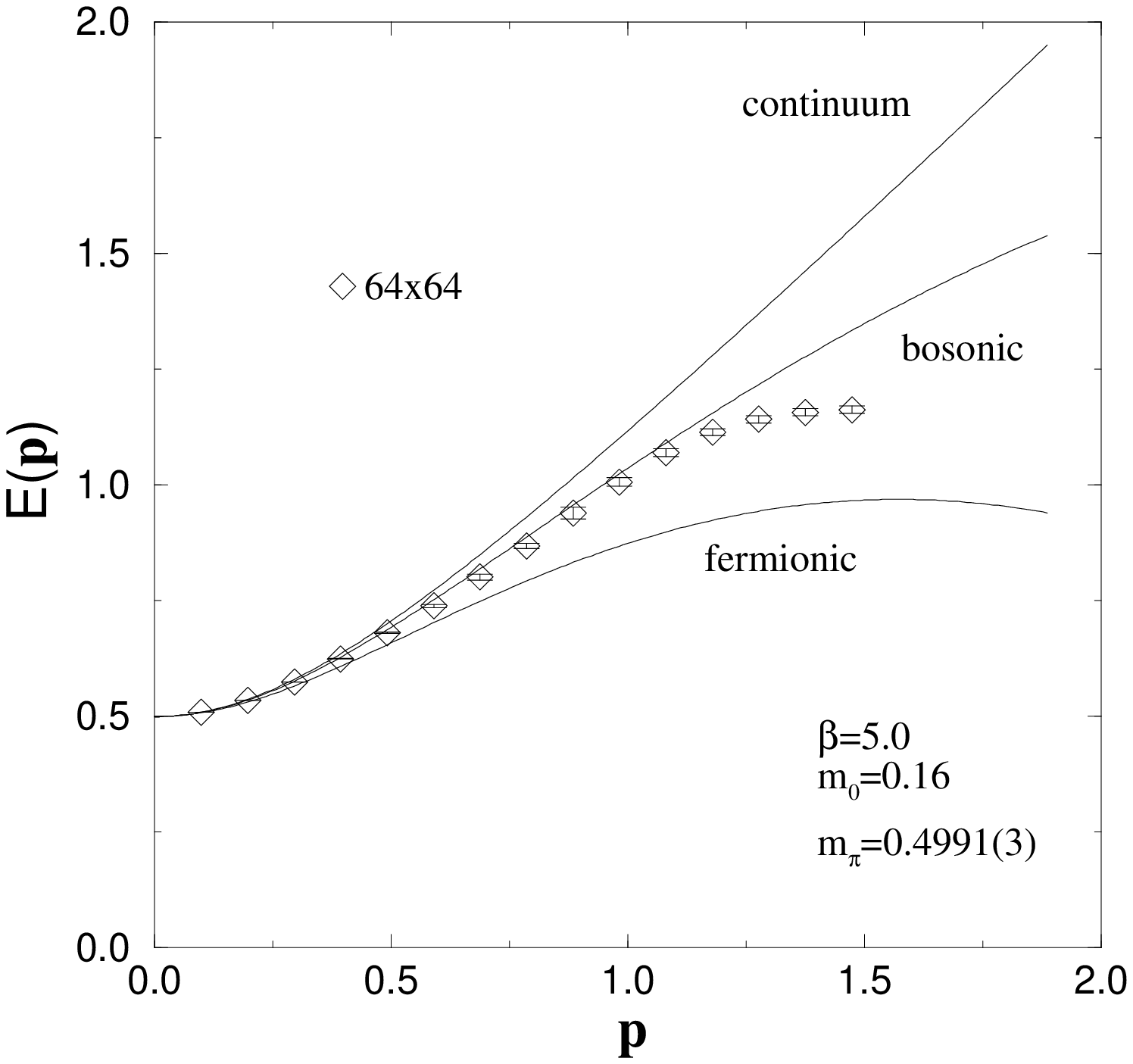,width=1.2\textwidth}
\emp
\begin{minipage}[b]{0.5\textwidth}
\hspace*{-0.6cm}\epsfig{file=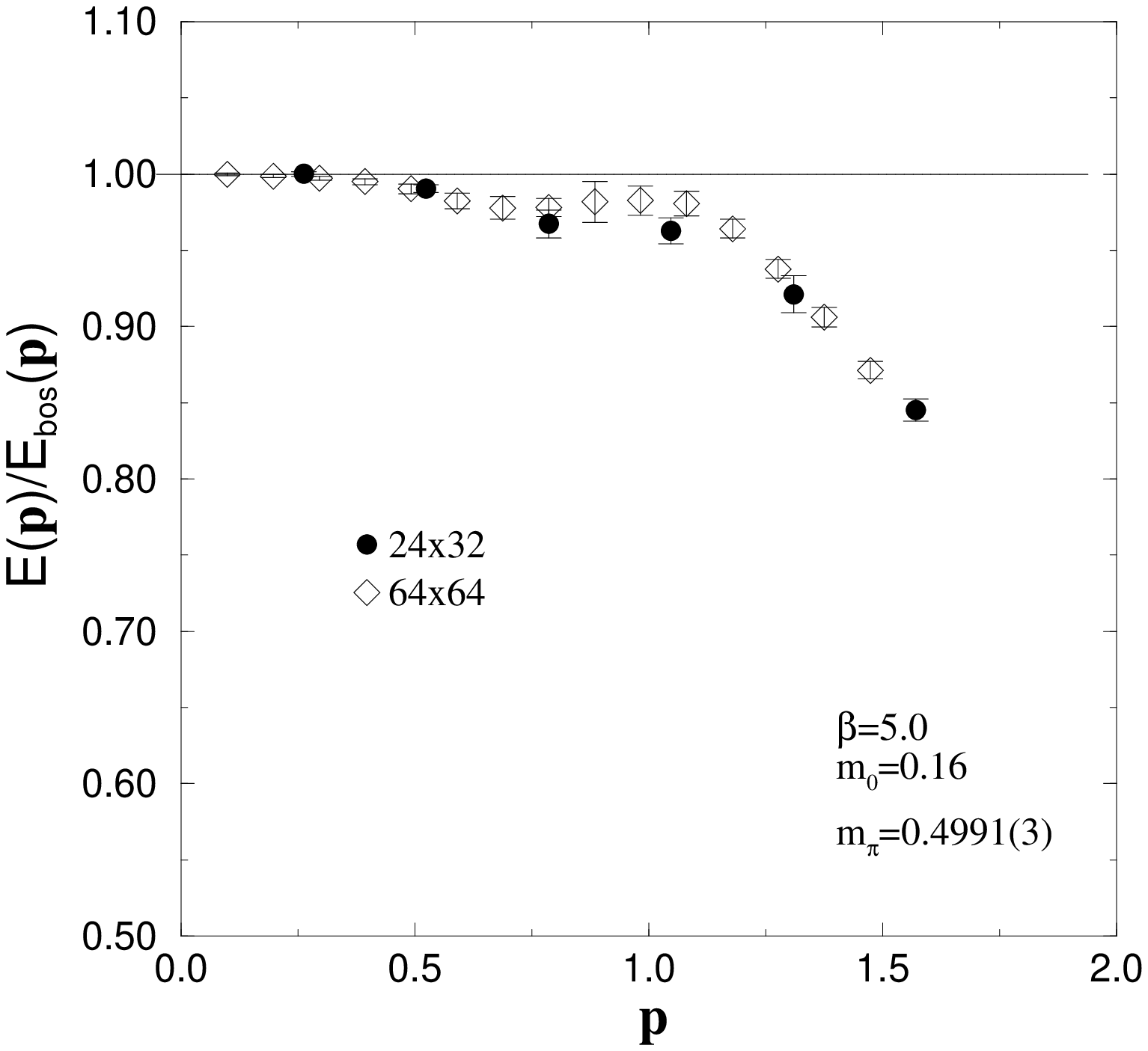,width=1.2\textwidth}
\emp\\
\capskip
\caption{\capsize Comparison of the different theoretical dispersion
  relations with the numerical results  for the pion energy
  (left). Ratio of the pion energy to the bosonic lattice dispersion
  relation for two different lattices (right).}
\label{massnum-fig4}
\end{figure}

For the determination of the particle momenta and the elastic scattering
phases of the pions a correct dispersion
relation has to be used to reduce $\ord{a}$--errors. 
The results for
the energy--momentum relation $E(\pb)$ of the pion in fig. \ref{massnum-fig4}
can be compared with the continuum and the lattice dispersion
relations. 
As expected the bosonic dispersion relation
\bea
  \la 2 \sinh \frac{E}{2} \ra^2 & = & \la 2 \sinh \frac{m}{2} \ra^2
    + \hat{\pb}^2 \co \comm{\hat{\pb} \cdef 2 \sin \frac{\pb}{2}}
\label{massnum-eq15}
\eea 
gives the best description for the numerical data.  

A better resolution 
of the difference between numerical results and the bosonic 
dispersion relation
is obtained by taking the quotient of the simulation
data and the bosonic lattice dispersion relation (second diagram in
fig. \ref{massnum-fig4}). 
Significant deviations from the theoretical
curve which are above one percent occur for both lattices for momenta
$\pb > 0.5$. This has to be compared with the simulation results
concerning the scattering phases in section
\ref{streunum-results}. The momenta of the scattering pions with masses
$m_\pi \le 0.4$ are in the region $\pb=0.1 \ldots 0.2$. Errors which
occur by using the bosonic dispersion relation are therefore negligible.

\subsect{Light mesons}
In the strongly coupled Schwinger model the (approximate) scaling of
the pion mass $m_\pi(m_0)$ is predicted (see section \ref{predict}).
With regard to the corresponding formula 
(\ref{schwinger-eq2}) we
check asymptotic scaling in the numerical simulations. Equation
(\ref{schwinger-eq2}) is also used to determine the strong
coupling region in which the predictions from the sine--Gordon model
are reliable. 

\begin{figure}[tb]
\vspace*{-1.5cm}
\hspace*{0.5cm}
\epsfig{file=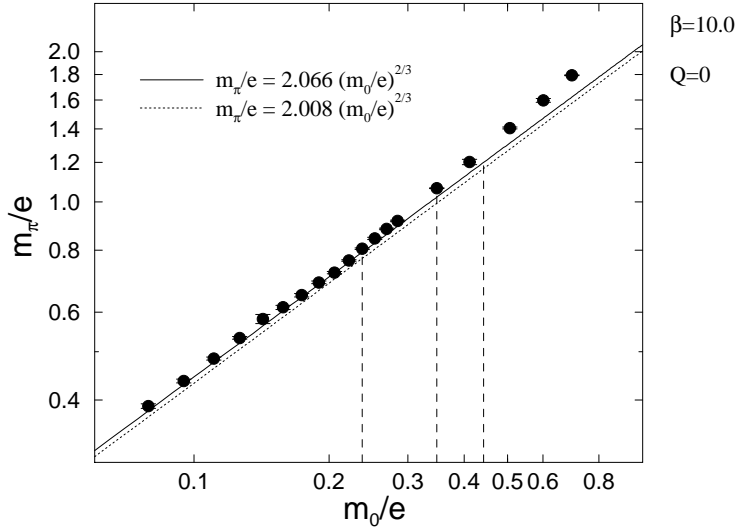,height=0.5\textheight}\\
\capskip
\caption{\capsize Comparison of the numerical results for $m_\pi(m_0)$
  ($L\times T = 48\times 48$ to $86\times 86$)  
  with the analytical predictions (\ref{schwinger-eq3}) and
  (\ref{schwinger-eq2}). The parameters marked with dashed
  lines are the simulation points that have been used for the
  determination of the scattering phases (see section \ref{streunum-results}).}
\label{massnum-fig1}
\end{figure}

First numerical results for the pion mass with reasonable
agreement
with (\ref{schwinger-eq3}) for $m_0/e \le 0.1$ were presented 
in ref. \cite{Ke86}. The results from ref. \cite[fig.1]{GK98} for
$\beta=4$ show
that a scaling of the pion mass according to 
\bea
  \frac{m_\pi}{e} \propto \la \frac{m_0}{e} \ra^p\,,\comm{p=0.689(10)}
\label{massnum-eq6}
\eea
exists for $m_0/e$ up to $0.4$. 
Nevertheless there are deviations for
$\beta=2$ and $\beta=4$ in ref. \cite[fig.1]{GK98} from the 
analytically predicted curve. 
As the continuum limit is reached for $\beta \rightarrow \infty$ 
these deviations decrease for higher values of the coupling
$\beta$ as expected. For $\beta=10$ and $Q=0$ the simulation data 
for the pion mass are compared with the analytical ones 
(\ref{schwinger-eq3}) and (\ref{schwinger-eq2})
in the double logarithmic plot in
fig. \ref{massnum-fig1}. 
The data
for $m_0/e < 0.25$ nearly coincide with the semiclassical
formula (\ref{schwinger-eq3}) and are slightly above the exact 
prediction (\ref{schwinger-eq2}).
From these small deviations between numerical data and analytical formulae
one can conclude that for couplings $\beta \ge 10$
the expectation value on the lattice is dominated by the topological
sector $Q=0$. It is therefore possible to neglect non--trivial
topological sectors for high values of $\beta$. 
In contrast to the region $m_0/e < 0.25$ it is visible in  
fig. \ref{massnum-fig1} that for $m_0/e \ge
0.4$ the pion mass does not scale very well
according to (\ref{schwinger-eq2}).

\begin{figure}[p]
\begin{minipage}[b]{0.5\textwidth}
\hspace*{-0.8cm}\epsfig{file=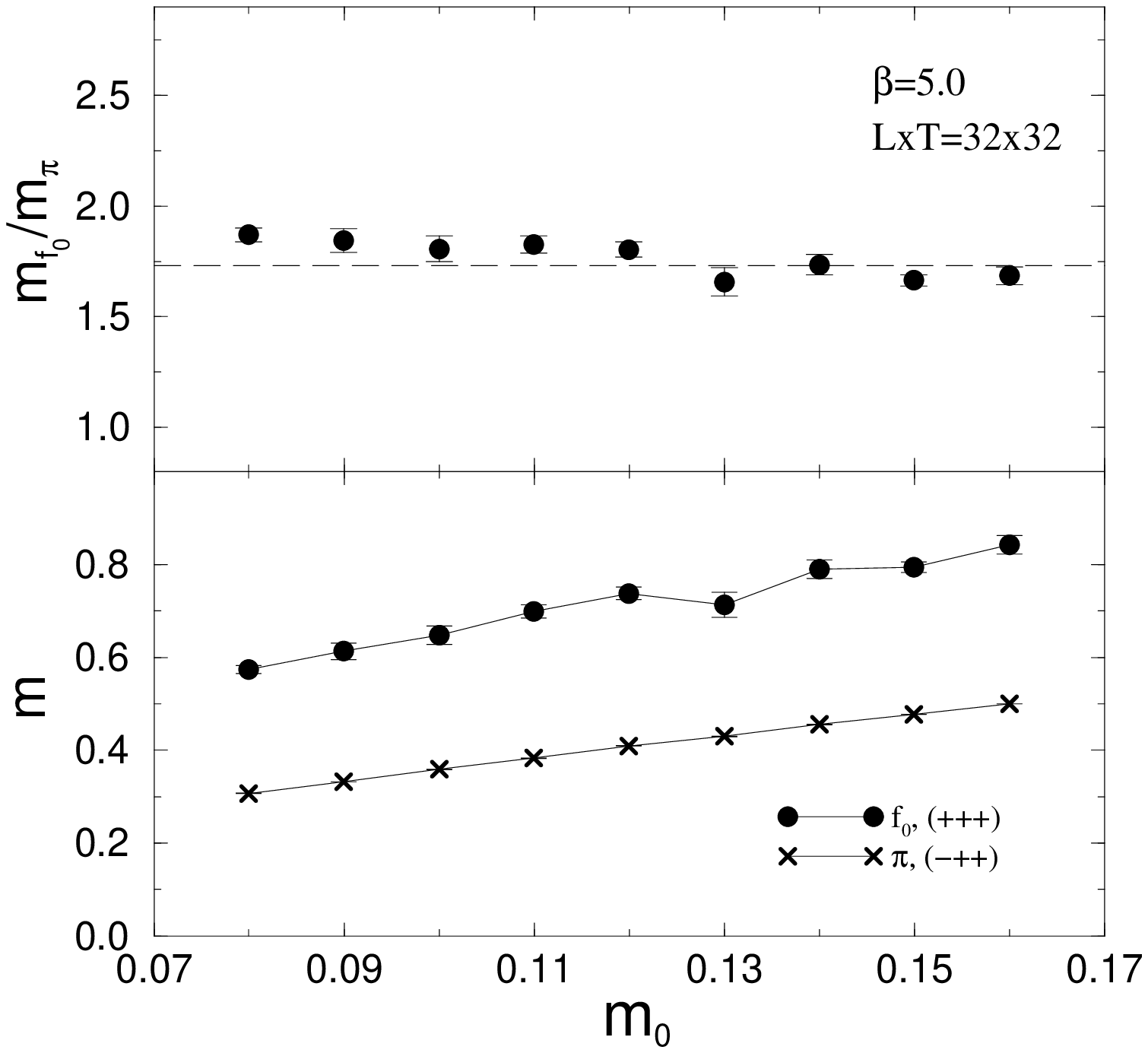,width=1.3\textwidth}
\emp
\begin{minipage}[b]{0.5\textwidth}
\hspace*{-0.6cm}\epsfig{file=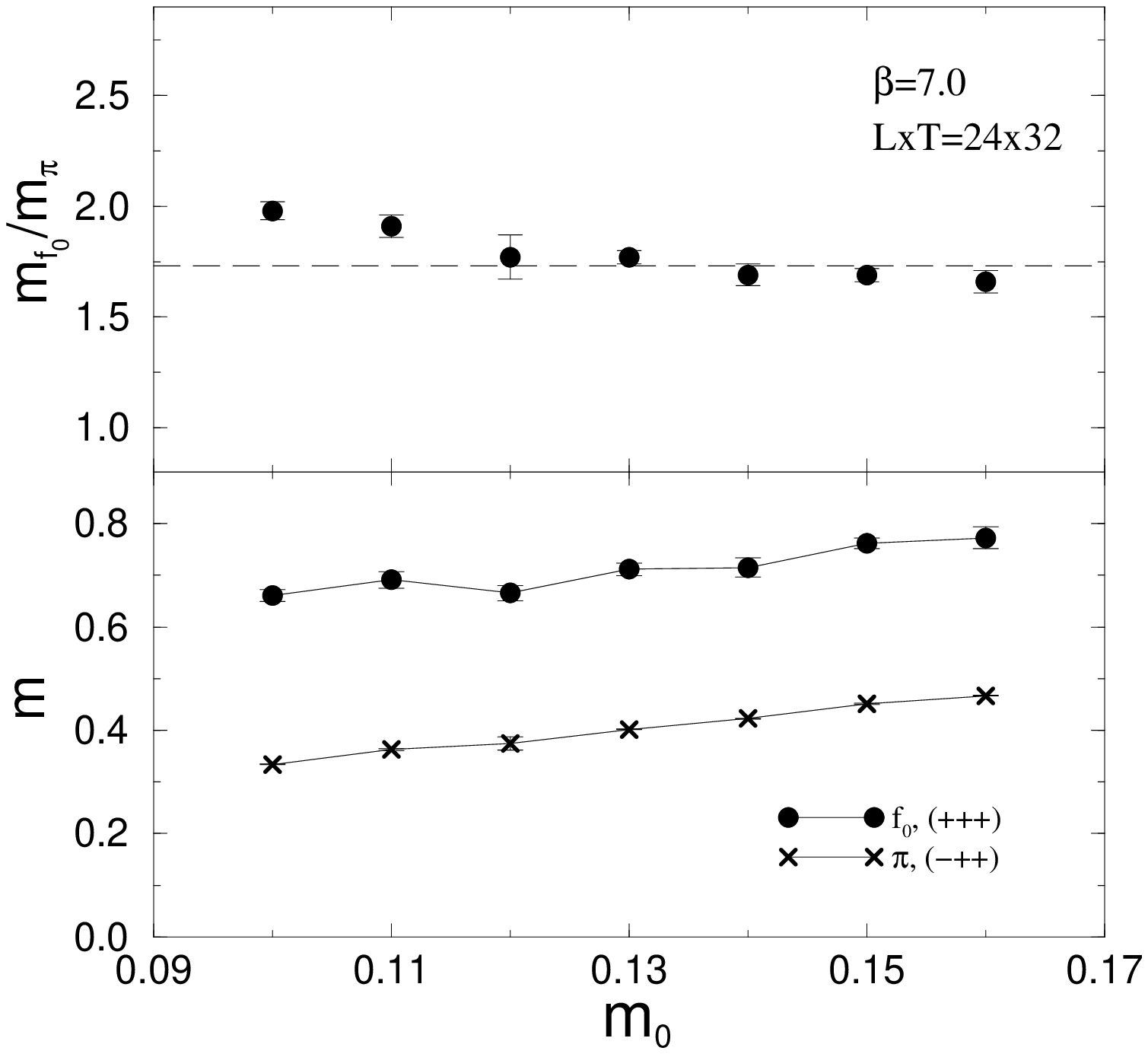,width=1.3\textwidth}
\emp\\
\capskip
\caption{\capsize Mass ratio $m_{f_0}/m_\pi$ for $\beta = 5.0$
and $\beta = 7.0$. The expected value for the mass ratio for $m_0/e
\rightarrow 0$ according to eq. (\ref{schwinger-eq4}) is 
$m_{f_0}/m_\pi = \sqrt{3}$ (dashed lines).}
\label{massnum-fig3}
\end{figure}

Besides the pion a scalar singlet with $I^{PG} = 0^{++}$
(``$f_0$''--meson) is predicted in 
the sine--Gordon model for $\beta_{SG} = \sqrt{2\pi}$. 
In the
simulations a singlet with the correct quantum numbers can be
identified. To compare the data with the predictions from section
\ref{predict} we used the dimensionless mass ratio $m_{f_0}/m_\pi$. 
This ratio is expected from eq. (\ref{schwinger-eq4}) to be 
$m_{f_0}/m_\pi = \sqrt{3}$. The results for the $f_0$ and pion mass for
$\beta=5$ and $\beta=7$ are
plotted in fig. \ref{massnum-fig3}. For a bare mass of about $m_0
\simeq 0.14$ the mass ratio from the simulations agrees with the
predicted value. 
But for smaller values of the bare mass there are
deviations from the $\sqrt{3}$--line in fig. \ref{massnum-fig3}. 

\begin{figure}[p]
\vspace*{-1.5cm}
\hspace*{0.7cm}
\epsfig{file=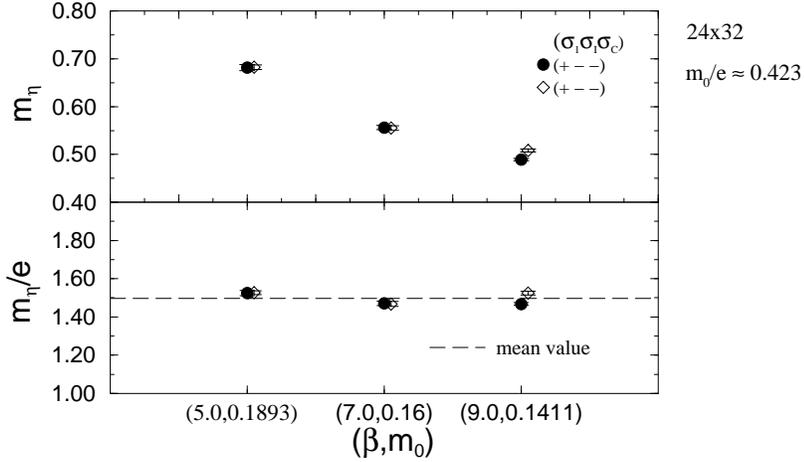,height=0.5\textheight}\\
\capskip
\caption{\capsize Mass of the $\eta$--meson for two different
  operators in the same lattice sector. Three different values of
  $\beta$ and $m_0$ are used with a fixed ratio $m_0/e \simeq 0.16 \sqrt{7}$.}
\label{massnum-fig7}
\end{figure}

\begin{figure}[tb]
\begin{minipage}[b]{0.5\textwidth}
\hspace*{-0.6cm}\epsfig{file=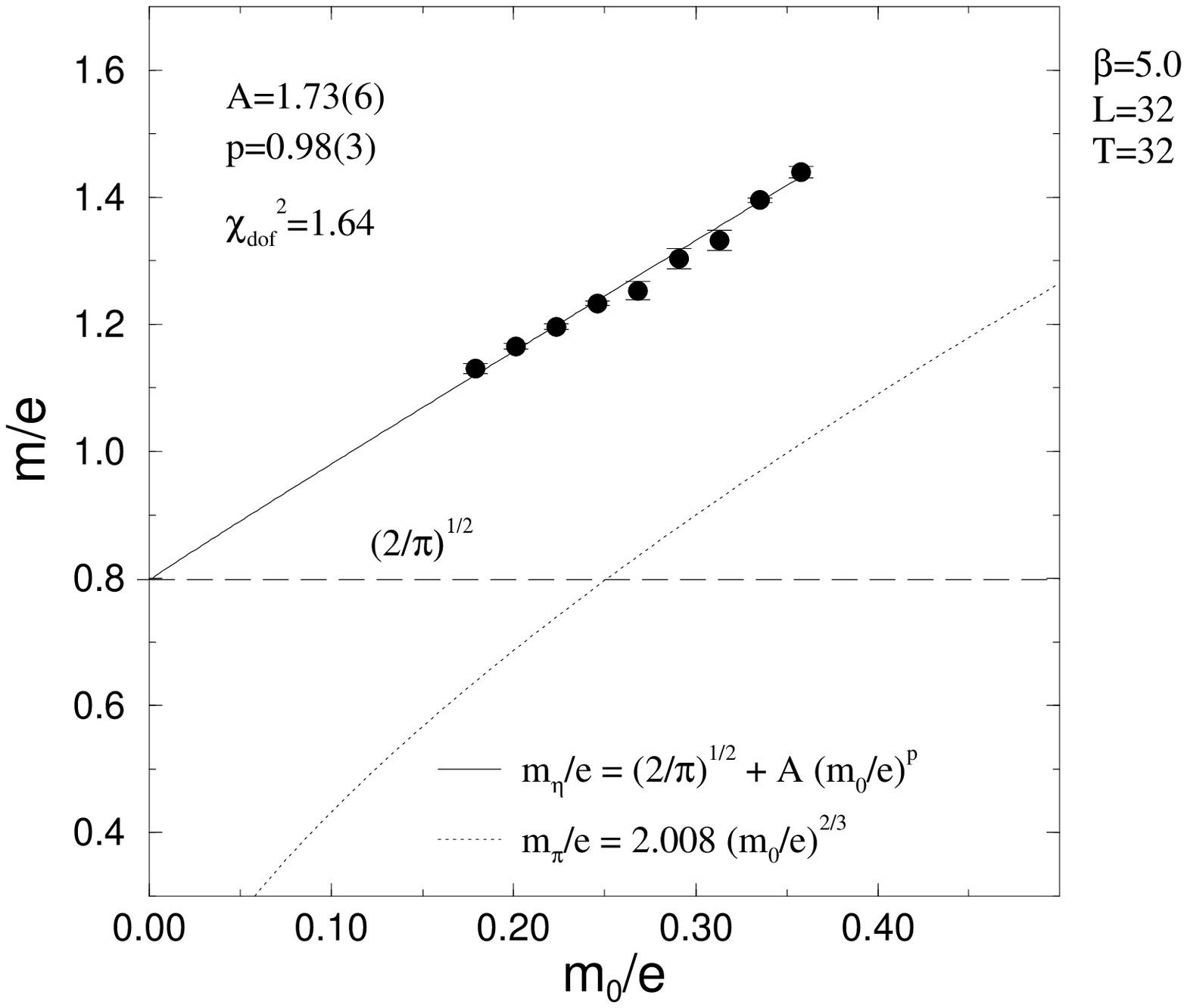,width=1.2\textwidth}
\emp
\hspace*{-0.2cm}
\begin{minipage}[b]{0.5\textwidth}
\hspace*{-0.6cm}\epsfig{file=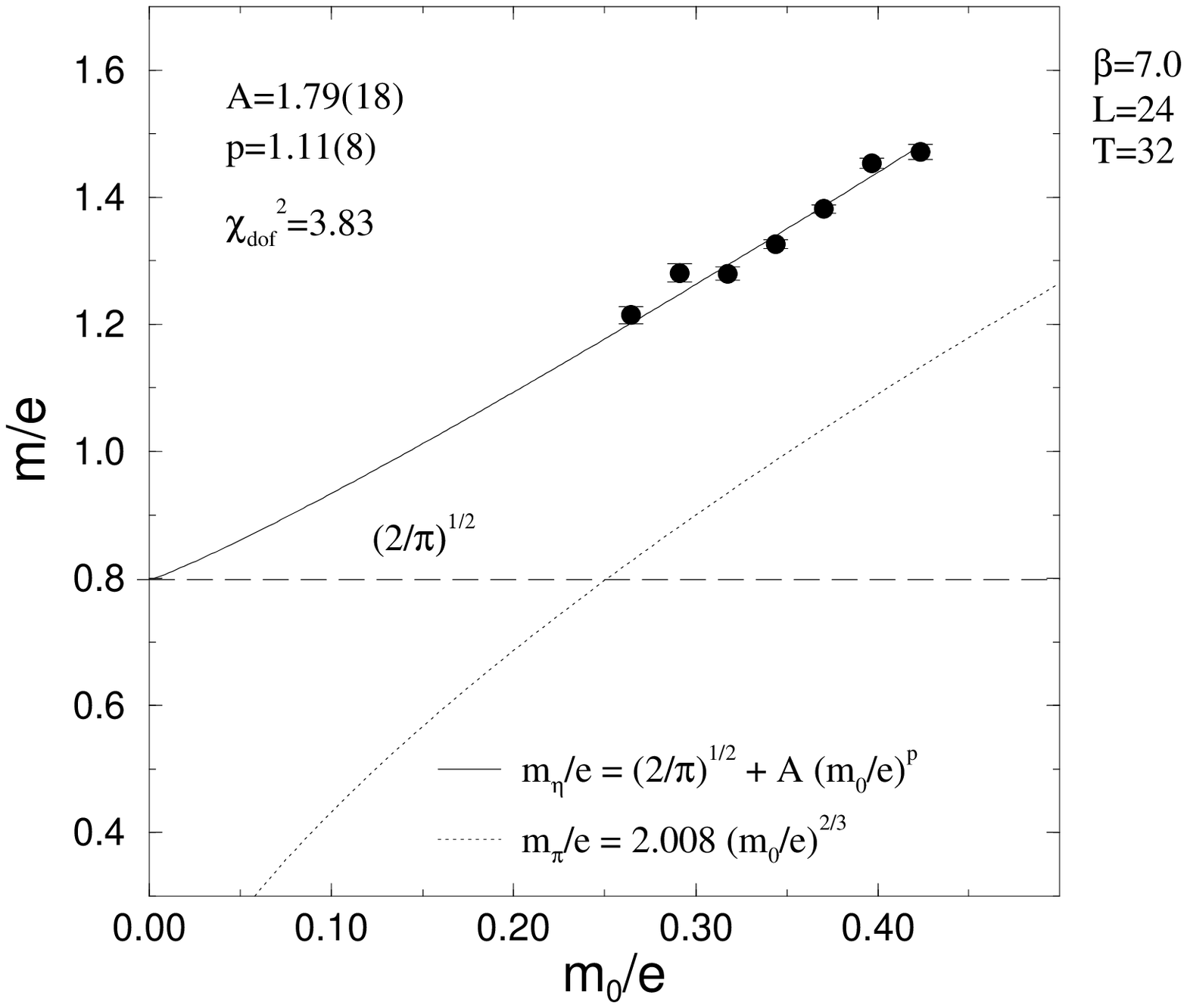,width=1.2\textwidth}
\emp\\
\capskip
\caption{\capsize Dependence of the $\eta$--mass on $m_0/e$ for
  $\beta=5$ and $\beta=7$. The ratio $m_\pi/e(m_0/e)$ according to
  formula (\ref{schwinger-eq2}) is plotted for comparison.}
\label{massnum-fig8}
\end{figure}

The third predicted particle in the Schwinger model is a pseudoscalar
singlet (``$\eta$''--meson). In the massless case $m_0=0$ its mass is
$m_\eta/e = \sqrt{2/\pi}$ (eq. (\ref{schwinger-eq5})). It follows
from dimensional arguments that
corrections to this value in the massive model should depend on
$m_0/e$ only. 
Fixing the ratio $m_0/e$ in lattice simulations it turns out
that for couplings $\beta \ge 5$ the ratio $m_\eta/e$ is indeed nearly
constant (see fig. \ref{massnum-fig7}).  
The simplest ansatz for the dependence of the $\eta$--mass on the
parameters in the Schwinger model 
is therefore:
\bea
\frac{m_\eta}{e} = \sqrt{\frac{2}{\pi}} + A \la \frac{m_0}{e}\ra^p \pu
\eea
This function can be used to fit the numerical results for $m_\eta$ 
for fixed coupling and various bare masses.
The results for $\beta=5$ and $\beta=7$
in fig. \ref{massnum-fig8} show that in both cases a nearly linear
dependence with $p \simeq 1$ and similar amplitudes $A$ is obtained.    
Because of the value of $p$ and $A$ the $\eta$--mass is larger than the
mass of the pion in the parameter region considered
($m_0/e < 0.5$).
This statement is also correct for the simulation points used for
the determination of the scattering phases. Hence the lowest
two--particle energies for these simulation points result from a
two--pion system.

Analogously to the pion (see section \ref{hmc}) 
we expect that the masses of the 
$f_0$-- and $\eta$--meson for different topological sectors 
are significantly different for large $\beta$.
Therefore the calculated masses
for different ensembles with a different probability 
distribution $p(Q)$ 
but same parameters might be different. We expect that the 
resulting errors for the data, especially for $\beta=7$ and $\beta=9$ in 
figs. \ref{massnum-fig3},\ref{massnum-fig7} and \ref{massnum-fig8}, 
are small due to the fact that all simulations were started
with $Q=0$ and hence the sector $Q=0$ strongly dominates in ensembles
for $\beta=7$ and $\beta=9$.

\begin{figure}[tb]
\begin{minipage}[b]{0.5\textwidth}
\hspace*{-0.5cm}\epsfig{file=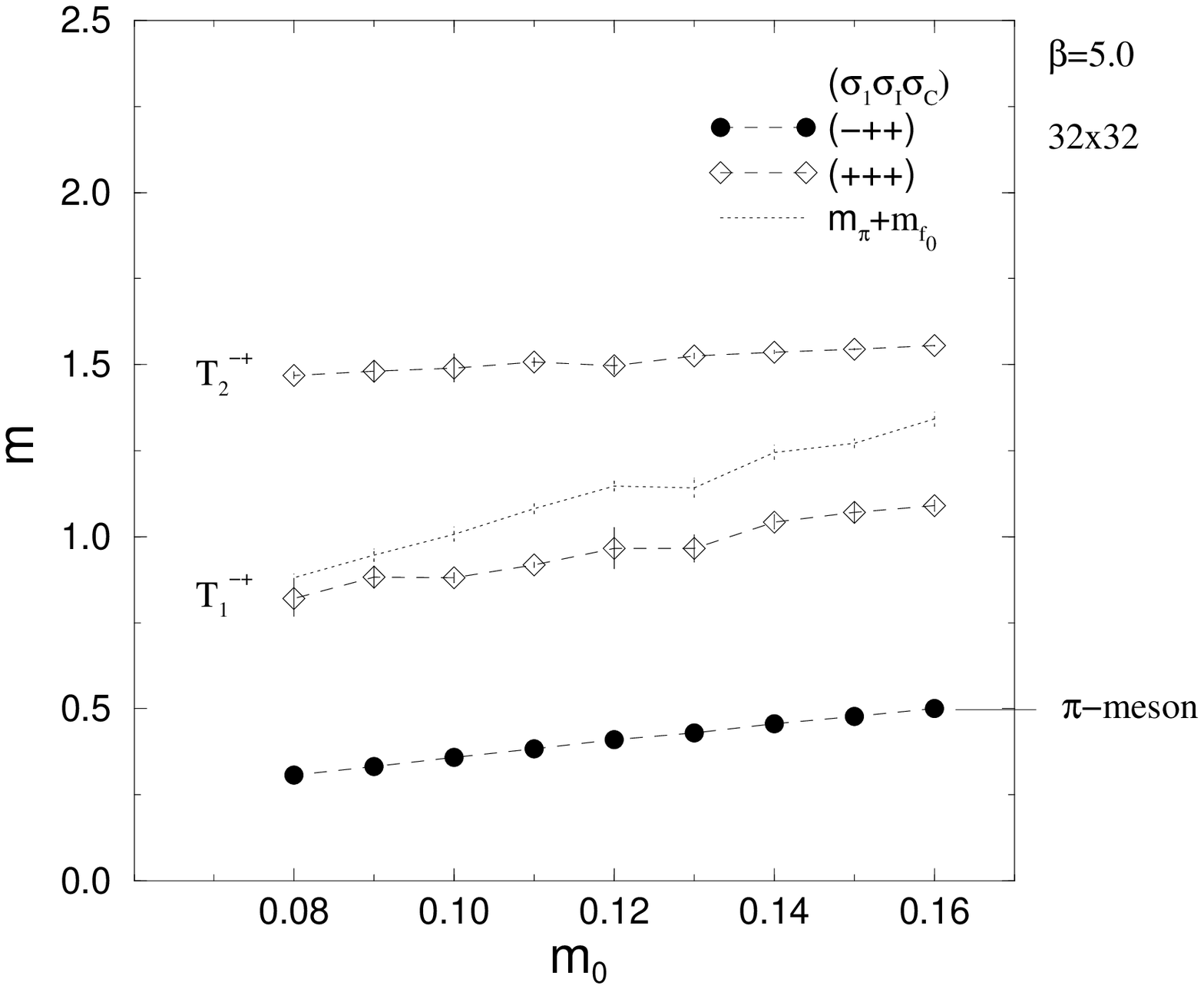,width=1.2\textwidth}
\emp
\hspace*{-0.5cm}
\begin{minipage}[b]{0.5\textwidth}
\epsfig{file=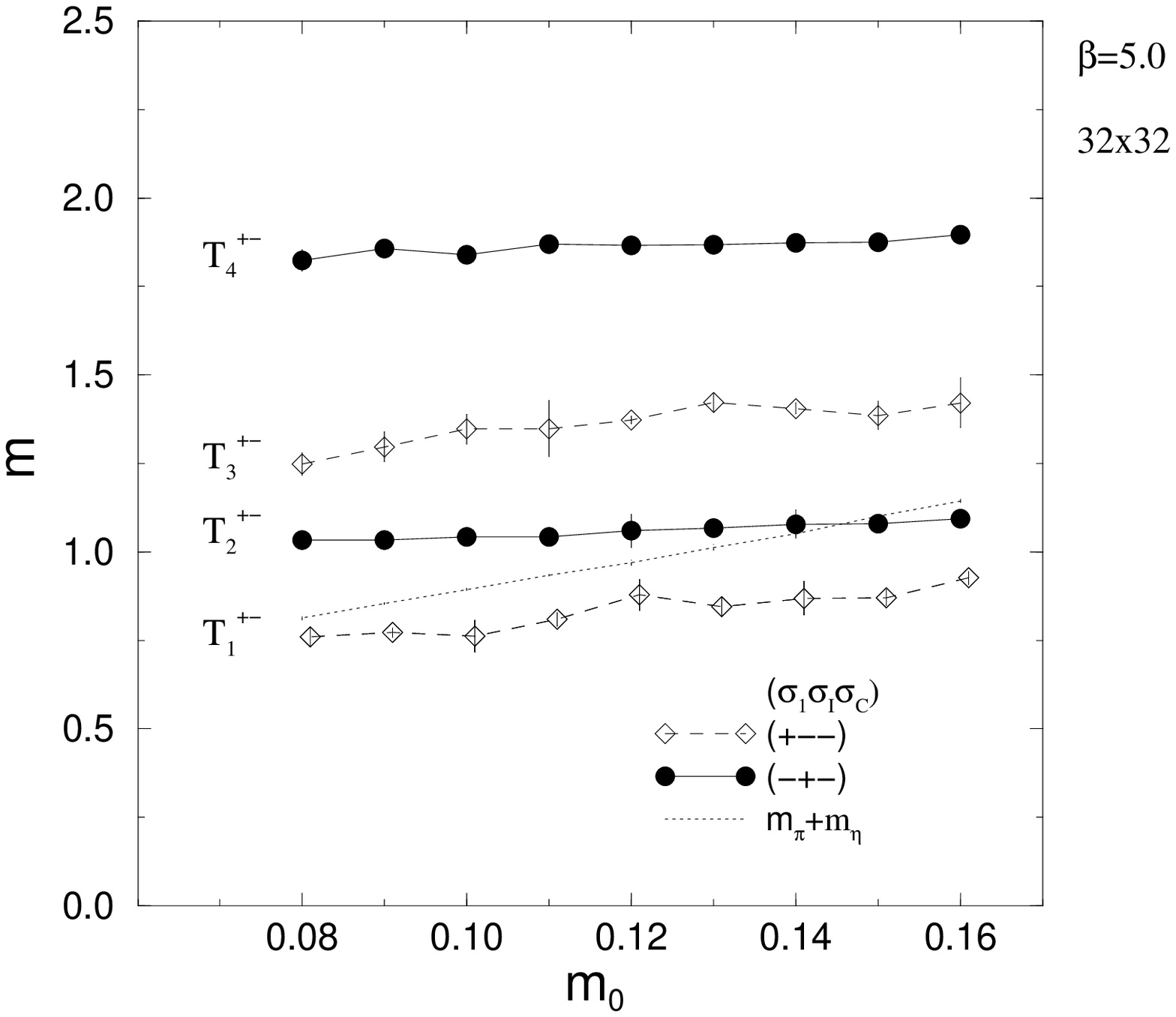,width=1.2\textwidth}
\emp\\
\capskip
\caption{\capsize Pseudoscalar triplets with $\sgg=+1$ 
  ($I^{PG} = 1^{-+} \equiv T^{-+}$) (left) and
  scalar triplets with $\sgg =-1$ ($I^{PG} = 1^{+-} \equiv T^{+-}$) (right). 
  The dotted lines represent
  the sum of the masses of those particles which might be possible
  decay products of the plotted states.}
\label{massnum-abb1}
\end{figure}

\begin{figure}[tb]
\begin{minipage}[b]{0.5\textwidth}
\hspace*{-0.5cm}\epsfig{file=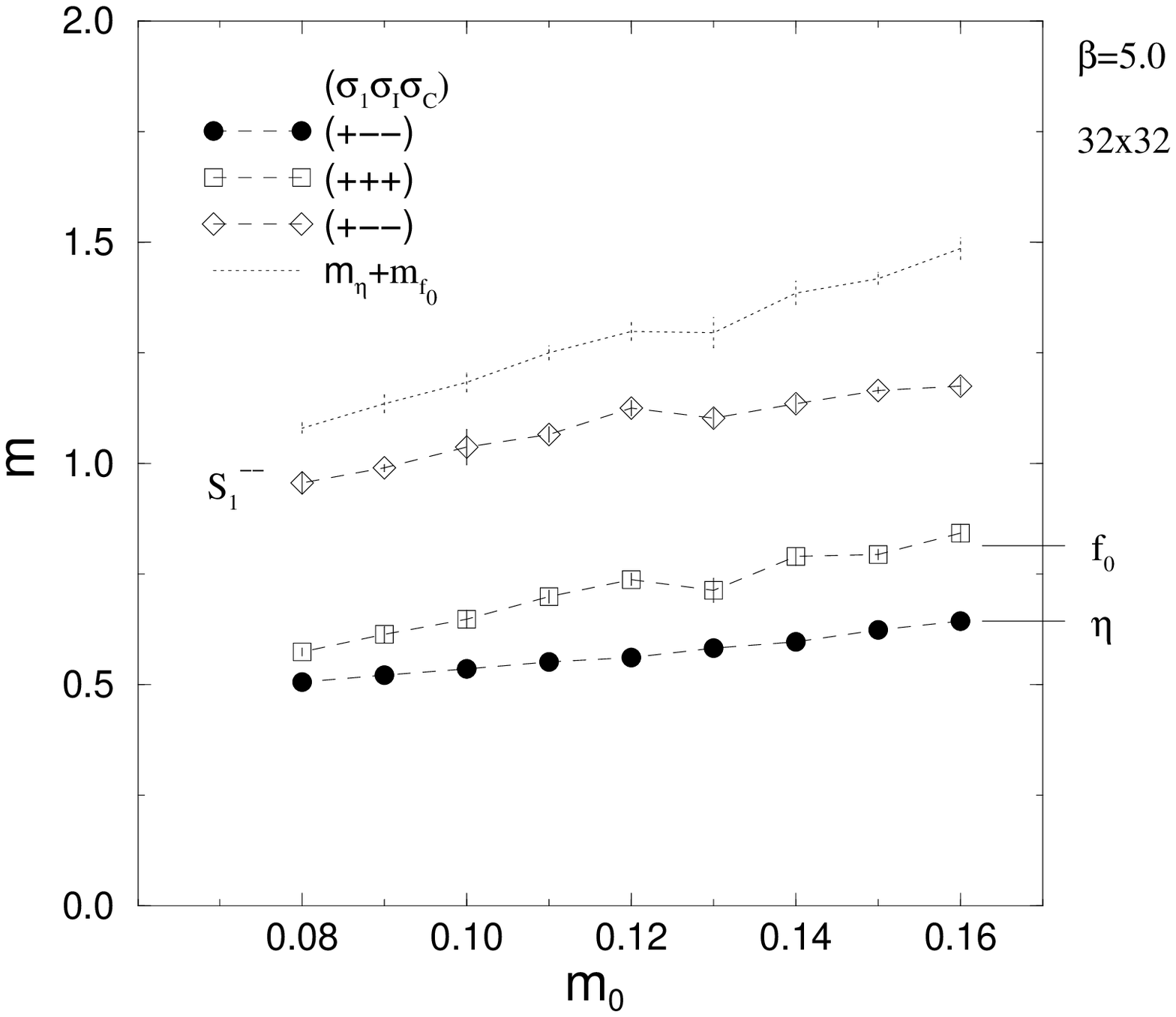,width=1.2\textwidth}
\emp
\hspace*{-0.4cm}
\begin{minipage}[b]{0.5\textwidth}
\epsfig{file=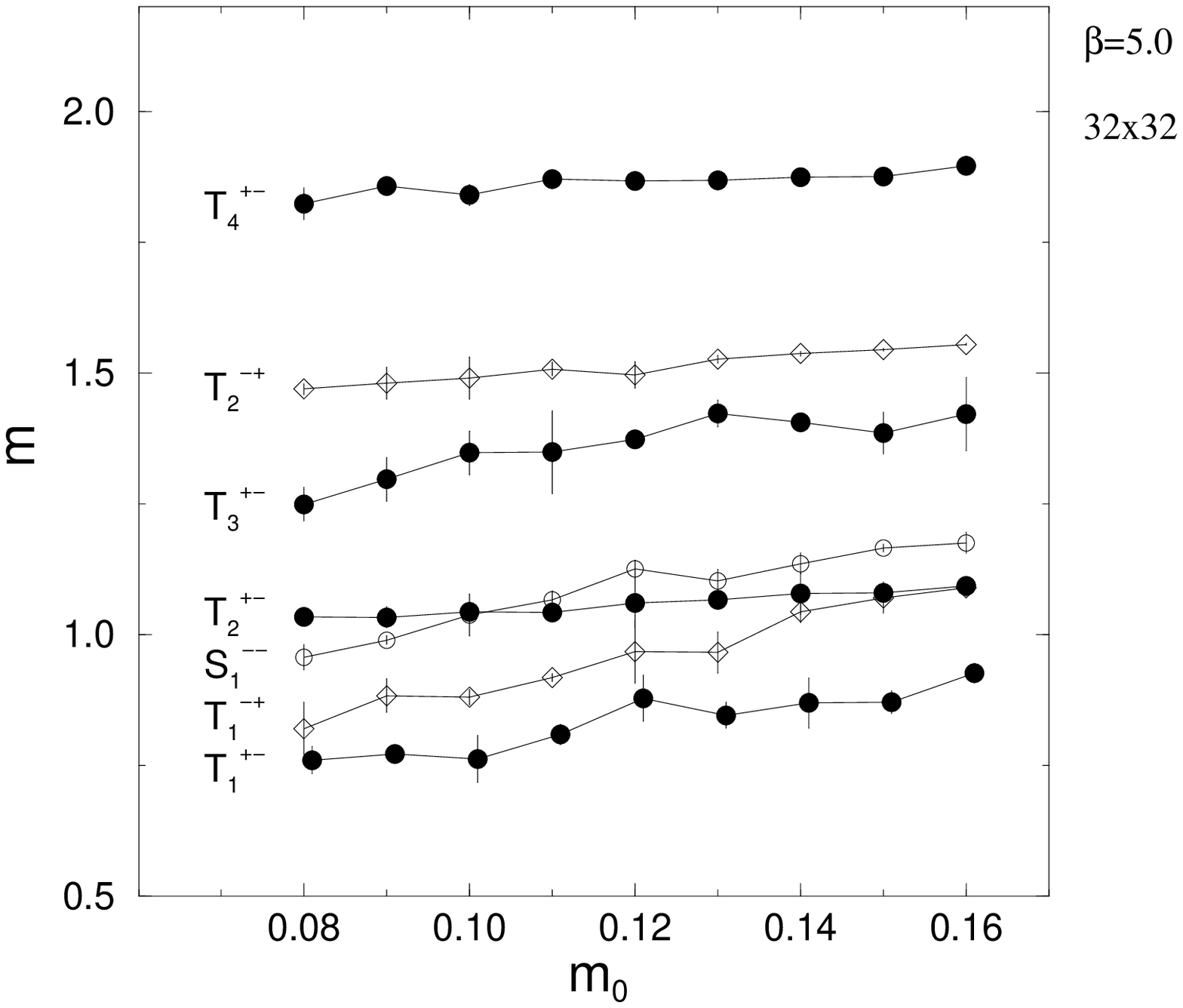,width=1.2\textwidth}
\emp\\
\capskip
\caption{\capsize Left: Different particles on the lattice which correspond
  to singlet states in the continuum ($I^{PG} = 0^{PG} \equiv
  S^{PG}$). Right: Complete overview of 
  the measured states apart from the three light mesons
  $\pi$, $f_0$ and $\eta$.}
\label{massnum-abb2}
\end{figure}

\subsect{Heavy Mesons} \label{otherstates}
In addition to the predicted particles in the Schwinger model 
we observe a large
number of other scalar and pseudoscalar particles. 
The lowest states which could be determined with sufficient accuracy
are shown in figs. \ref{massnum-abb1} and \ref{massnum-abb2}. 
The masses of these states are above the masses of the three
predicted mesons. 

Many of these states might be unstable. For example, it follows from
fig. \ref{massnum-abb1} that the
$T_2^{PG} = T_2^{-+}$--particle might decay into a 
$\pi$-- and a $f_0$--meson. The
states which are definitely stable in the examined parameter region 
are the $T_1^{-+}$--, $T_1^{+-}$-- and the 
$S_1^{--}$--meson. That means that these particles do not have decay
channels as far as the determined mesonic states are concerned.
There might be other particles like 
triplets with quantum numbers ($\sgp = +1$, $\sgg = +1$) which
have only a slight overlap with the operators we used in our simulations. 

In figs. \ref{massnum-abb1} and \ref{massnum-abb2}
masses with $m > 1$ are to be understood mainly qualitatively,
because one expects substantial $\ord{a}$--corrections. 

Summarizing the results for the mass spectrum we 
see that we are able to identify
the predicted mesons ---
$\pi$, $f_0$ and $\eta$ --- and their masses
in Monte--Carlo simulations. The analytical predictions
for the scaling of the pion-- and $f_0$--mass (see
section \ref{predict}) are confirmed with small deviations. Best results
are obtained for $\beta=10$ whereas for $\beta \le 5$
deviations from the continuum results occur. 
Considering the measurements for the various couplings 
the predictions from the sine--Gordon model 
for the strong coupling region of the Schwinger model 
are reliable for $m_0/e < 0.25$. 

Concerning the $\pi$--$\pi$ scattering those particles with $m \simeq
2m_\pi$ are of interest because they
might occur as resonances in the elastic scattering processes
or they might be confused
with the two--pion--energies in finite volume
as it will be shown in section \ref{massspectrum}.

\sect{Scattering phases} \label{streunum}
The aim of this section is to present the methods and results for the
determination of the scattering phases for the elastic
$\pi$--$\pi$ scattering in the Schwinger model.  
As already mentioned in the introduction we use
a formula which was
derived by L\"uscher for massive quantum field theories
\cite{Lu86a,Lu86b,LW90,Lu91a}. This method is based on the
fact that for not too small volumes the energies of the
two--particle states in finite volume are determined by the elastic
S--matrix of the model in infinite volume. Therefore L\"uscher's method
is suitable for the investigation of scattering processes on the
lattice with Monte--Carlo simulations. 

\subsect{L\"uscher's method} \label{streunum-luescher}
The system we are interested in consists of
two bosonic particles of mass $m$ in $1+1$ dimensions 
in a finite box with spatial length $L$. 
We work in the center--of--mass system with zero total momentum\footnote{An
extension of L\"uscher's method to systems with non--zero total
momentum was proposed and successfully applied in ref. \cite{GR95}.}.  
The formula for the determination of the scattering phases
$\delta$ for the elastic scattering of the particles in {\em
infinite} volume was given in ref. \cite{LW90}:
\bea
  2 \delta(\kb) = - \kb\,L \com{mod $2\pi$} \pu \label{streunum-eq1}
\eea 
$\kb$ is the momentum of one particle in the finite box 
in the center--of--mass system. It is related to the discrete
two--particle energy $E$ by
the relativistic formula:
\bea
  E = 2 \sqrt{m^2 + \kb^2} \pu
\eea
For a potential with short range it can be easily shown that equation
(\ref{streunum-eq1}) is exact in quantum mechanics \cite{LW90}.
Also for a bosonic scalar quantum field theory in four dimensions a
relation between the scattering phases and the particle momenta in finite
volume can be derived \cite{Lu91a}. 

An exact derivation of (\ref{streunum-eq1}) in an arbitrary quantum
field theory is in general not possible. Nevertheless it is plausible
that (\ref{streunum-eq1}) holds
provided that \cite{LW90}
\begin{itemize}
\item the examined quantum field theory
        has no massless particles in its spectrum
\item the mixing of the relevant two--particle states with other
  many--particle states is excluded
\item the range of the two--particle interaction is much smaller than
  half the spatial extension.
\end{itemize}
In the sine--Gordon model no inelastic scattering processes exist
\cite{ZZ79}. Nevertheless such inelastic processes are possible in 
the Schwinger model. Therefore in the
Schwinger model one has to ensure
that the lowest two--particle energies are below the inelastic
threshold $E=4m$. Furthermore,
one has to guarantee
that no production of other particles occurs in the relevant
energy region.  

If the extension of the spatial volume is too small, further
exponentially decaying finite size effects (polarization effects) appear 
in the two--particle energies. This leads to  
corrections to formula (\ref{streunum-eq1}). As far as the
scattering phases are concerned it is difficult to determine
numerically these finite size effects. 
To reduce these unwanted effects on the two--particle energies 
the minimum lattice size for all simulations is chosen
so that the finite size effect for the pion mass is
negligible. 

\subsect{Correlation functions} \label{streucorr}
For the determination of the two--particle energies in the finite
volume we start with the successful ansatz of refs. \cite{LW90,GK94a,GK96}. 
This means that we construct a correlation matrix with
time--slice--operators where different operators are characterized by
different orthogonal functions:
\bea
  O_i^{LT_\tau}(t) & = & \frac{1}{L^2} \sum_{\xb,\yb} 
  \cphi{\xb,\yb} \cb{\xb,t} \cc{\xb,t}\,
  \cb{\yb,t+\tau} \cc{\yb,t+\tau} \, \om_i(\yb-\xb)\co \label{streunum-eq4} \\
  & & \comm{\cphi{\xb,\yb} = (\pm 1)^\xb (\pm 1)^\yb,\,\, \tau=0,1\,,} \nn \\ 
  & & \comm{\om_i(\xb) = \lga \ba{l}\cos(\pb_i \xb)\\ \sin(\pb_i \xb) \,,
      \shspace \pb_i  =  \frac{2\pi}{L} i,\;\; i = 0,1,\ldots \pu \ea \right.}\nn 
\eea
These operators having the property that the fermion and antifermion in
each meson operator are local in time are denoted by 
$LT\equiv LT_0$ ($\tau=0$) and $LT_1$ ($\tau=+1$).  
The definition of the functions $\om_i(\xb)$ and $\cphi{\xb,\yb}$ 
depends on the choice of the lattice symmetry sector. 
In some sectors it is also possible
to use two definitions for $\cphi{\xb,\yb}$: 
$\cphi{\xb,\yb} \equiv (-\sge)^\xb (-1)^\yb$ or 
$\cphi{\xb,\yb} \equiv \sge^\xb$. 
The corresponding operators are denoted with a 
$1$ and $2$ respectively appended to the quantum numbers.

The operator (\ref{streunum-eq4}) has the quantum number $\sgc=+1$. 
To construct eigenstates on the lattice which are odd under charge
conjugation we use
\bea
  O_i^{NLT}(t) = \frac{1}{L^2} \sum_{\xb,\yb} 
  \cphi{\xb,\yb} \,\cb{\xb,t+1} \cc{\xb,t}\,
  \cb{\yb,t} \cc{\yb,t+1} \,\cum{2}{\xb,t} \cumast{2}{\yb,t}
  \om_i(\yb-\xb) \co \label{streunum-eq2}
\eea 
which in general is not an eigenstate under charge conjugation.
This is an operator where the fermion and antifermion in each meson
are non--local in time ($NLT$). Like in the case of the $LT$-operators the
definition of $\om_i(\yb-\xb)$ and $\cphi{\xb,\yb}$ is determined by
the symmetry sector.

The advantage of the special 
construction (\ref{streunum-eq2}) is visible if one constructs
an operator with a definite quantum number $\sgc$:
\bea
{O_{\sgc,\,i}^{NLT}} (t) & \equiv & O_i^{NLT}(t) + \sgc \la \ltsC O_i^{NLT} \ra (t) \nn\\
               & = & \sum_{\xb,\yb} \cb{\xb,t+1} \cc{\xb,t} \,\cb{\yb,t}
               \cc{\yb,t+1} \,\cum{2}{\xb,t} \cumast{2}{\yb,t}
               \om_i(\yb-\xb) \nn\\ 
            & & \hspace{5cm} \mal \lga \cphi{\xb,\yb} +
               \sgc\sgi\cphi{\yb,\xb} \rga \pu \label{streunum-eq3}
\eea
Because of the expression 
(\ref{streunum-eq3}) the calculation of Green--functions with
the $NLT$--operator is not more expensive than the calculation
with the $LT$--operator.

As it will be shown in the following section the four--meson 
correlation functions of the operators (\ref{streunum-eq4}) and 
(\ref{streunum-eq2}) include contributions from one--particle
states. 
To avoid such contributions for the determination of the 
$\pi$--$\pi$ scattering phases it appears advisable to
calculate correlation functions which are
constructed according to the $(I=2)$--sector in the continuum
because there no one--particle states occur.
Details will be seen in section \ref{massspectrum}.
The procedure of the derivation 
of these correlation functions is the same as for the $\cpi$--function:
The starting--point here are the expectation values 
$C_{cont,ij}(t) = \langle O_{i}(t) \tilde{O}_{j}(t) \rangle $
in the continuum. 
The operators $O_i$ and $\tilde{O_j}$ are defined by
\bea
O_{i}(t) & \equiv & \frac{1}{(2\pi)^2} \int d\xb d\yb \,
           \om_i(\yb-\xb)
           \psib^1(\xb,t) \gamma_5 \psi^2(\xb,t) \,
           \psib^1(\yb,t) \gamma_5 \psi^2(\yb,t) \,,\nn \\
\tilde{O}_{j}(t) & \equiv & \frac{1}{(2\pi)^2} \int d\xb d\yb \, 
        \om_j(\yb-\xb)
        \psib^2(\xb,t) \gamma_5 \psi^1(\xb,t) \,
        \psib^2(\yb,t) \gamma_5 \psi^1(\yb,t) \pu \nn
\eea 
By naive translation of the correlation functions $C_{cont,ij}(t)$ 
into the lattice formulation one obtains among others
the following terms
\bea
  \lefteqn{ C_{\pi\pi,ij}(t) } \nn\\
   & \equiv & \frac{\sgi}{L^3} \sum_{\xb,\yb,\zb} {\langle}\, 
      | \Mi{{\bf 1},{\bf 3}} |^2 \, | \Mi{{\bf 2},{\bf 4}} |^2 
    + | \Mi{{\bf 2},{\bf 3}} |^2 \, | \Mi{{\bf 1},{\bf 4}} |^2 \nn\\
  & &  - 2 \mbox{Re} \lba \Micc{{\bf 1},{\bf 4}} \, \Mi{{\bf 1},{\bf 3}} \, 
           \Micc{{\bf 2},{\bf 3}} \, \Mi{{\bf 2},{\bf 4}} \rba \,{\rangle}_U \nn\\[0.1cm]
  & &  \mal (-1)^{{\bf 1}+{\bf 2}+{\bf 3}+{\bf 4}} \, \om_i(\yb-\xb)
    \om_j(\zb) \, \cphi{\xb,\yb} \cphi{\zb,0}\,, \label{streuop-gl5} \\[0.2cm]
  & &  \com{${\bf 1} \equiv (\xb,t),\hspace{1cm} {\bf 2} \equiv (\yb,t+\tau)$},   \nn\\
  & &  \com{${\bf 3} \equiv (\zb,0),\hspace{1cm} {\bf 4} \equiv (0,\tau^\prime)$}\nn\pu  
\eea     
The prefactors in (\ref{streuop-gl5})
are chosen in such a way that the expression 
(\ref{streuop-gl5}) coincides with corresponding terms in the
correlation function of the operators (\ref{streunum-eq4}).
In the following the index $\mbox{}_{\pi\pi}$ always denotes a
correlation function of the type (\ref{streuop-gl5}).
For $\tau=\tau^\prime=0,1$ the expression (\ref{streuop-gl5}) is related
to the $LT$,$LT_1$--operator and for $\tau=1$, $\tau^\prime=0$ to an
$LT_1$--$LT$ correlation function. 
If one takes the naive continuum limit of these lattice correlation functions
it turns out that only the expression (\ref{streuop-gl5}) contributes
to the expectation values $C_{cont,ij}(t)$. Hence we expect
the $C_{\pi\pi}(t)$--function to yield the dominant contribution to
the $I=2$ continuum sector and therefore to be sufficient to calculate
the two--particle energies in this sector.

The advantage of the expression (\ref{streuop-gl5}) is obvious. 
Compared to the calculation of e.g. 
$\langle O^{LT}(t) O^{LT}(0) \rangle$ the calculation of
(\ref{streuop-gl5}) needs merely a fraction of the conjugate
gradient calculations.

\begin{figure}[tb]
\vspace*{-1.5cm}
\epsfig{file=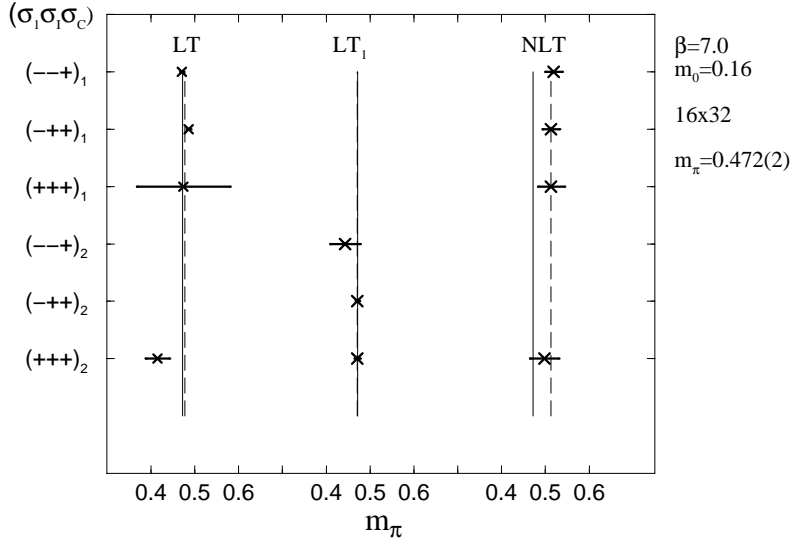,height=0.5\textheight}\\
\capskip
\caption{\capsize Results for the pion mass for different
  four--meson correlation functions in different lattice sectors.
  The dashed line is a fit to the data, 
  whereas the solid line represents the
  mass obtained from the two--pion correlation function $\cpi$.}
\label{streunum-pic1}
\end{figure}

\subsect{One--particle states in four--meson correlation functions} \label{massspectrum}
The numerical results for the four--meson correlation functions with
the operators (\ref{streunum-eq4}) and (\ref{streunum-eq2}) show
that in nearly all sectors the main
contributions and hence the lowest energies 
can be identified with the single--particle masses of the model. 
Just as for 
the two--meson correlation functions 
the pion--state here also occurs with the
best signal--to--noise ratio. The comparison of the determined masses
with the pion--mass calculated with the 
two--pion correlation function $\cpi(t)$
shows a good correspondence
between the results for the two--meson and four--meson
correlation functions (see fig. \ref{streunum-pic1}).  

This result
can be explained in terms of the general properties of the 
$\tsu$ flavour group of the Schwinger model:
The structure of the group $\tsu$  is such that some irreducible
representations may be expressed
either in terms of tensors of rank four or tensors of
rank two. Therefore it is
possible to examine properties of particles with isospin $1$,
i.e. triplets, with fermionic eight--point functions. According to
table \ref{einb} the lattice sectors which are connected to a  
$I^{PG}=1^{-+}$ continuum state
are $(\sge,\sgi,\sgc)$ = $(--+)$, $(-++)$ and $(+++)$.
In fact we observe the pion in all these 
sectors (see fig. \ref{streunum-pic1}). 
Besides the pion we have determined also other mesons --- the $f_0$-- and
$T_1^{+-}$--meson ---
with appropriate
fermionic eight--point functions.

According to table \ref{einb} the pion as well as the
$\pi$--$\pi$ states occur in most sectors with
$\sgc=+1$. Therefore the lowest energy and
consequently (presumably) also the signal--to--noise ratio in these
sectors is dominated by the pion. Furthermore, according to
the results of section \ref{otherstates} there exist 
pseudoscalar mesons with masses $m \simeq 2 m_\pi$. These particles
possibly could make the determination of the
two--particle energies more difficult. 
To avoid these problems we used for the reasons given above
the $C_{\pi\pi}(t)$--function (\ref{streuop-gl5})
for the determination of the scattering phases.
In fact the numerical results
discussed in section \ref{streunum-results} show
that by using the $C_{\pi\pi}(t)$--function the contributions
of the single--particle states are suppressed. 
In particular
the lowest energies in the $C_{\pi\pi}(t)$--function are
the $\pi$--$\pi$ energies.

\subsect{Connected correlation functions} \label{connected}
One might suspect that the only  
cause for the appearance of the one--particle energies in some
four--meson correlation functions are poles 
in the disconnected parts of the correlation functions. 
To refute this hypothesis we examined a 
four--meson correlation function $C_{ij}(t) = \lla O_i^{LT}(t) O_j^{LT}(0)
\rra$.
(The following discussion also holds for other operators ($LT_1$,
$NLT$, \ldots ) with more general space--time
arguments of the fermions.)    
This correlation function separates into a sum of connected
correlation functions and disconnected parts:
\begin{fmffile}{feynfmf}
\be
C_{ij}(t) =  
\raisebox{-17\ul}{
\begin{fmfgraph*}(40,40)
  \fmfsurround{e1,e2,e3,e4,e5,e6,e7,e8}
  \fmfv{d.sh=circle,d.fi=.5,d.si=.5h}{v1}
  \fmf{plain}{v1,e2}
  \fmf{plain}{v1,e4}
  \fmf{plain}{v1,e6}
  \fmf{plain}{v1,e8}
\end{fmfgraph*} \
} \
 + \mbox{disconnected graphs} \pu
\ee 
The shaded symbol represents the connected Green--function 
($\langle \ldots \rangle_c$).
The external legs denote
quark--antiquark pairs in which the incoming mesons on the left
live on the time slice $t$ and the outgoing mesons on the right
on the time slice $0$.

The disconnected parts consist of 
\bea
  \raisebox{-17\ul}{                
\begin{fmfgraph*}(40,40)
  \fmfsurround{e1,e2,e3,e4,e5,e6,e7,e8}
  \fmfv{d.sh=circle,d.fi=.5,d.si=.3h}{v1}
  \fmfv{d.sh=circle,d.fi=.5,d.si=.3h}{v2}  
  \fmf{plain}{v1,e4} 
  \fmf{plain}{v1,e6}
  \fmf{plain}{v2,e2}
  \fmf{plain}{v2,e8}
  \fmf{phantom}{v1,v2}
\end{fmfgraph*}}_{ ({\bf A})} \
+ \
\raisebox{-17\ul}{                
\begin{fmfgraph*}(40,40)
  \fmfsurround{e1,e2,e3,e4,e5,e6,e7,e8}
  \fmfv{d.sh=circle,d.fi=.5,d.si=.3h}{v1}
  \fmfv{d.sh=circle,d.fi=.5,d.si=.3h}{v2}  
  \fmf{plain}{v1,e2} 
  \fmf{plain}{v1,e4}
  \fmf{plain}{v2,e6}
  \fmf{plain}{v2,e8}
  \fmf{phantom}{v1,v2}
\end{fmfgraph*}}_{ ({\bf B})} \
+ \
\raisebox{-12\ul}{                
\begin{fmfgraph*}(30,30)
  \fmfsurround{e1,e2,e3,e4,e5,e6,e7,e8}
  \fmfv{d.sh=circle,d.fi=.5,d.si=.4h}{v1}
  \fmf{plain}{v1,e4}
  \fmf{plain}{v1,e8}
\end{fmfgraph*}} 
\raisebox{-12\ul}{                
\begin{fmfgraph*}(30,30)
  \fmfsurround{e1,e2,e3,e4,e5,e6,e7,e8}
  \fmfv{d.sh=circle,d.fi=.5,d.si=.4h}{v1}
  \fmf{plain}{v1,e2}
  \fmf{plain}{v1,e6}
\end{fmfgraph*}}_{ ({\bf C})} 
 \label{streunum-eq5}
\eea
and of graphs like 
\be
\raisebox{-17\ul}{
\begin{fmfgraph*}(40,40)
  \fmfsurround{e1,e2,e3,e4,e5,e6,e7,e8}
  \fmfv{d.sh=circle,d.fi=0,d.si=.5h}{v1}
  \fmf{phantom}{v1,e2}
  \fmf{phantom}{v1,e4}
  \fmf{plain}{v1,e6}
  \fmf{phantom}{v1,e8}
\end{fmfgraph*}}
\mal \,\ldots \pu \label{streunum-eq6}
\ee
The dots in (\ref{streunum-eq6}) represent graphs with three external legs 
like 
\raisebox{-10\ul}{
\begin{fmfgraph*}(30,30)
  \fmfsurround{e1,e2,e3,e4,e5,e6,e7,e8}
  \fmfv{d.sh=circle,d.fi=.5,d.si=.5h}{v1}
  \fmf{plain}{v1,e2}
  \fmf{plain}{v1,e4}
  \fmf{phantom}{v1,e6}
  \fmf{plain}{v1,e8}
\end{fmfgraph*}}
\pu

In the usual procedure only the vacuum parts $({\bf A})$ are
subtracted from
the Green--function to obtain the connected Green--function
\bea
 \lla O_i(t) O_j(0) \rra_c \simeq \lla O_i(t) O_j(0) \rra -  \lla O_i(t)
 \rra \lla O_j(0) \rra \pu \label{streunum-eq15}
\eea
In our simulations it turns out that
the expressions $({\bf B})$ and $({\bf C})$ are not negligible and
must also be subtracted in eq. (\ref{streunum-eq15}).
However it can be derived easily that $({\bf B})$ as well as 
$({\bf C})$ disappear for
$i \ne j$ because of the orthogonality of the functions $\om_i(\yb-\xb)$.
For example, the term $({\bf B})$:
\bea
({\bf B}) & \propto & \sum_{\xb,\yb,\xbp,\ybp} \cphi{\xb,\yb} \om_i(\yb-\xb) 
 \om_j(\ybp-\xbp) \nn\\
& & \mal \lla \cb{\xb,t}\cc{\xb,t}\, \cb{\xbp,t_0}
   \cc{\xbp,t_0} \rra \lla \cb{\yb,t} \cc{\yb,t} \,\cb{\ybp,t_0} 
      \cc{\ybp,t_0}  \rra
\eea
can be rewritten to obtain 
\bea
({\bf B}) & \propto & 
\delta_{ij} \frac{L^2}{2} 
\sum_{\xb} \cphie{\xb} \cos(\pb_i \xb) 
\lla \cb{\xb,t} \cc{\xb,t} \,\cb{0,t_0} \cc{0,t_0} \rra \nn\\
& & \mal \sum_{\yb} \cphiz{\yb} \cos(\pb_i \yb) 
\lla \cb{\yb,t} \cc{\yb,t} \,\cb{0,t_0} \cc{0,t_0} \rra \co 
\label{streunum-eq7} \\
& & \comm{\cphi{\xb,\yb} \cdefr\, \cphie{\xb} \cphiz{\yb}} \co
    \comm{ i \ne 0,\frac{L}{2} } \pu \nn
\eea
\end{fmffile}%
In the same way one obtains $({\bf C}) \propto \delta_{i j}$. This 
means that the disconnected parts $({\bf B})$ and $({\bf C})$ 
contribute to the diagonal elements of the correlation matrix only.

This result shows that the one--particle contributions to the
four--meson correlation functions can not come from the
terms in (\ref{streunum-eq5}) only, because we obtain in
our simulations that the pion state
also occurs in the non--diagonal elements of the
correlation matrices (see table \ref{streunum-tab2}). 
\begin{table}[t]
\bea \la 
\ba{llll} 0.503(14) & 0.506(6) & 0.492(6) & 0.498(17) \\
          0.505(4)  & 0.50(2)  & 0.504(12)& 0.516(6)  \\   
          0.498(2)  & 0.50(4)  & 0.495(9) & 0.523(9)  \\
          0.494(2)  & 0.51(4)  & 0.495(9) & 0.526(9)  
\ea
\ra \nn \eea
\caption{\capsize Pion--masses which result from each of the $16$
  matrix elements for a $4\times 4$ correlation matrix. The values are
  obtained with an $LT$ operator for $\beta=5.0$, $m_0=0.16$ and
  $L\times T=24\times 64$ in the sector $(\sge,\sgi,\sgc)$ =
  $(--+)_2$. For comparison: The
  pion--mass for a single--particle operator is $m_\pi = 0.4991(3)$.}
\label{streunum-tab2}
\end{table}

Actually, the disconnected parts are a nuisance
for the determination of two--meson energies:
From equation (\ref{streunum-eq7}) it follows 
that the disconnected parts are 
products of propagators of moving mesons. Therefore from 
(\ref{streunum-eq7}) one obtains energies which are twice the
energy of a single free meson with momentum $\pb_i$.
In fact we could determine these energies with a good signal
in the Green--function $C_{ij}(t)$. The overall
results for $C_{ij}(t)$ as well as the measured
energies from the disconnected parts $({\bf B}) + ({\bf C})$ alone
are listed in table \ref{streunum-tab1}.
Comparing the energies which correspond to each
other from both measurements good agreement
within the errors is visible. 

These energies from the disconnected parts make it more difficult to
analyse the simulation data and to determine
two--meson energies we are mainly interested in. 
In some cases they might even be
confused with the two--pion energies. In order to avoid 
contributions from the expressions $({\bf B})$ and $({\bf C})$ 
we considered only correlations with $i \ne j$, because as we have seen
above all non--diagonal elements of $({\bf B})$ and $({\bf C})$
vanish. 
In fact we calculated $r\times r$
correlation matrices $D(t)$ which are not symmetric in their indices:
\bea
D_{ij}(t) & \cdef & C_{i,r+j}(t) \nn\\
             & = & \lla O_i(t) O_{r+j}(0) \rra, \comm{i,j=1\ldots r}
             \pu
\label{streunum-eq8}
\eea
Moreover, we did not use momenta $\pb_i =0,\pi$ for the wave functions. 
In this case it can be shown that the contributions from 
(\ref{streunum-eq6}) vanish, too.

Altogether we use the function $D_{\pi\pi}(t)$ for the calculation of 
the two--pion energies. This is
a correlation matrix (\ref{streunum-eq8}), for which according
to (\ref{streuop-gl5}), 
only those parts contribute which are relevant 
for the $\pi$--$\pi$ scattering in the $(I=2)$--sector. 
\begin{table}[t]
\renewcommand{\arraystretch}{1.2}
\begin{center}
\begin{tabular}{|l||>{$}l<{$}|>{$}l<{$}|>{$}l<{$}|}
  \hline
                &  C_{ij}(t)& (\bf B) + (\bf C) & 2 E_{bos}(m_\pi,\pb_i) \\\hline\hline
  $m_\pi$       &  0.506(5) & \multicolumn{1}{c|}{-----} & \multicolumn{1}{c|}{-----} \\\hline
  Energy $E_2$ &  1.10(3)  & 1.21(4) & 1.124(9) \\\hline
  Energy $E_3$ &  1.53(11) & 1.49(3) & 1.418(7) \\\hline
  Energy $E_4$ &  1.89(16) & 1.89(4) & 1.776(5) \\\hline
\end{tabular}
\caption{\capsize 
Comparison of the results from the four--meson correlation matrix
$C_{ij}(t)$ with the values resulting from the disconnected parts
$(\bf B) + (\bf C)$. The energies were obtained for the same
simulation parameters $\beta=5.0$, $m_0=0.16$, $L \times T = 24\times
64$ from a $4\times 4$ correlation matrix. Additionally the energies
which were calculated with the bosonic dispersion relation are listed ($m_\pi =
0.506(5)$, $\pb_i = 2\pi i/L$, $i=1\ldots 3$). 
}
\label{streunum-tab1}
\end{center}
\end{table}

Analyzing the numerical data for $D_{\pi\pi}(t)$ it turns out that
the best results are not obtained by calculating the generalized 
eigenvalues but just by diagonalizing the correlation matrix. 
In this case the eigenvalues show the following behaviour 
($T \rightarrow \infty$):
\bea
\lambda_l(t) \stackrel{t \rightarrow \infty}{=} c_l (\sigma_l)^t
\,\ce{-t E_l} \lba 1
+ \ord{\ce{-t\Delta E_l}} \rba \,, \shspace \sigma_l= \pm 1 \pu 
\label{streunum-eq9}
\eea
Like in eq. (\ref{methods-eq10}) $\Delta E_l$ 
denotes the smallest difference between $E_l$ and
other spectral energies. Following closely the 
proof in ref. \cite{LW90}, equation (\ref{streunum-eq9}) can be derived
easily for the $D_{\pi\pi}(t)$--correlation matrix.

\subsect{Results for the scattering phases} \label{streunum-results}
For the determination of the two--particle energies we calculated
a $4\times 4$ $D_{\pi\pi}(t)$--matrix with the $LT$--, $LT_1$--operators
(see section \ref{streucorr})
and the $LT_1$--$LT$ operator combination for all relevant
lattice sectors. The simulation results show that a clear signal for
the $\pi$--$\pi$ energies occurs only for the
$LT$--operator in the $(\sge,\sgi,\sgc)$ = $(+++)_2$ lattice sector. 
For the calculation of the energies in this case it was frequently
necessary to enlarge the correlation matrix to obtain a better
separation of the different energies. Hence we
extracted the lowest energy only
whereas higher energies could not be determined
with sufficient accuracy. 

To improve the signal for the lowest energy we use smeared
operators (see section \ref{masscorr}). 
The optimal value for the smearing parameter $\alpha$ 
from simulations with low statistics turned out to be
 $\alpha \simeq 0.02$. 
Here one has to be aware of the fact that the replacement of the
antifermions in a two--meson operator $O(t)$ with quantum number
$\sgc(O)$ by smeared antifermions (``smeared source'')
leads to a smeared operator $O^S(t)$ which has no
definite behaviour under charge conjugation. In such cases it is
useful to chose $\alpha$ so small to have dominant contributions from
the sector with $\sgc = \sgc(O)$.

\begin{figure}[tb]
\begin{minipage}[b]{0.45\textwidth}
\hspace*{-0.2cm}\epsfig{file=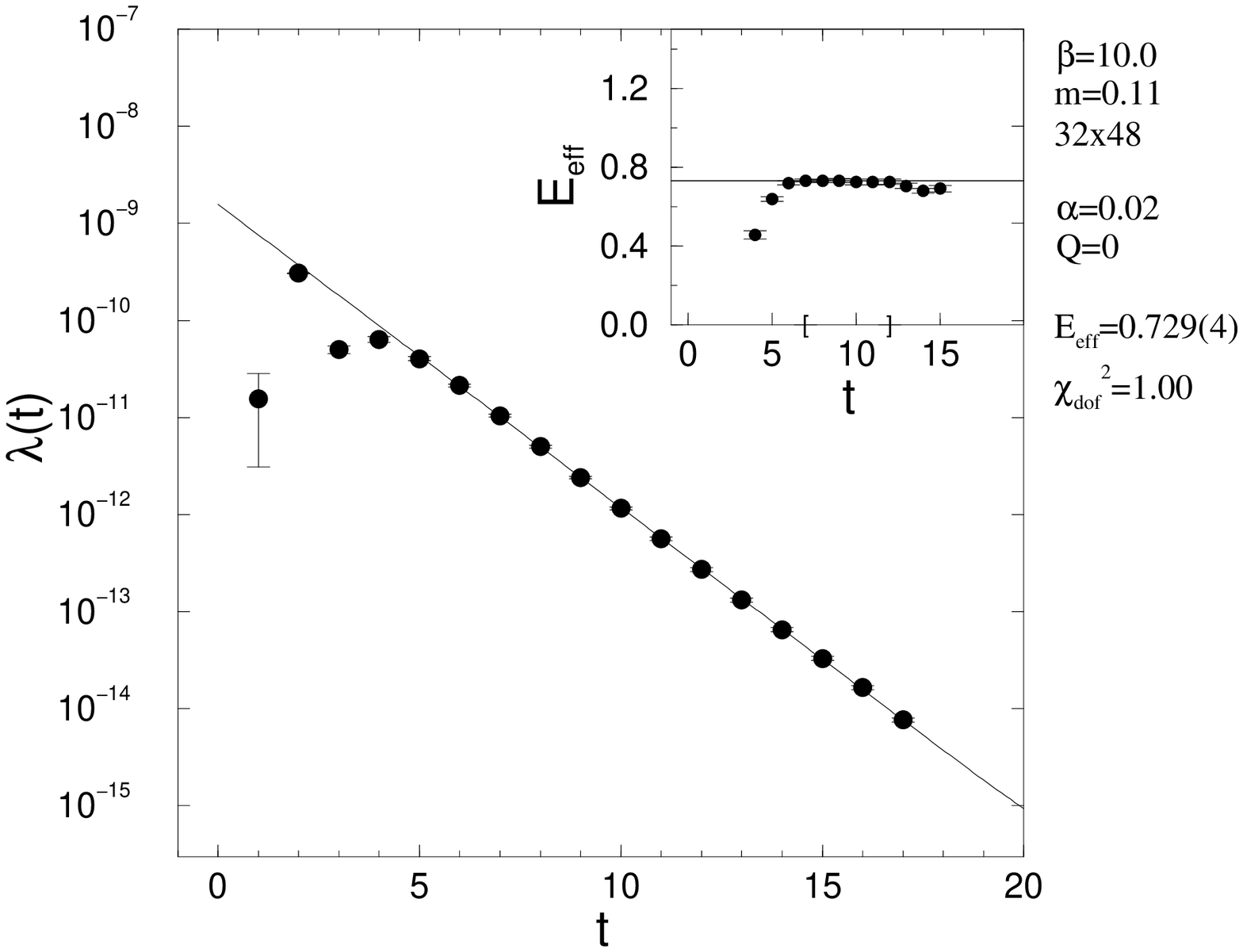,width=1.2\textwidth}
\emp
\begin{minipage}[b]{0.45\textwidth}
\hspace*{0.4cm}\epsfig{file=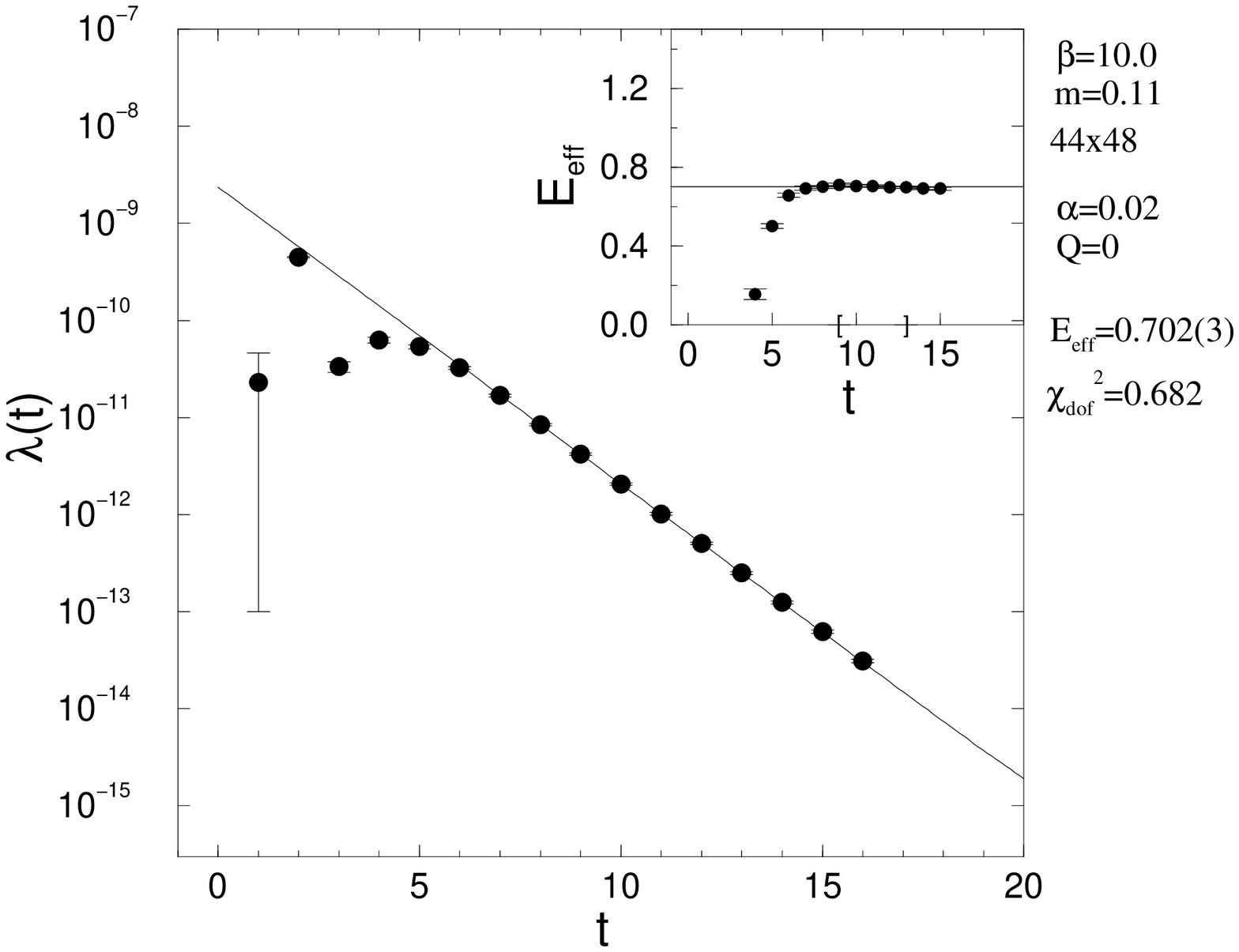,width=1.2\textwidth}
\emp\\
\capskip
\caption{\capsize Two--particle energies for ($\beta=10$, $m_0=0.11$ ) for 
two different spatial extensions of the lattice in the sector 
$(\sge,\sgi,\sgc)=(+++)_2$.
The lowest eigenvalue of the
correlation matrix and its effective energy are plotted.}
\label{streunum-fig3} 
\end{figure}

The results for the lowest eigenvalues of two
typical correlation matrices are shown in fig. \ref{streunum-fig3}. 
Sufficiently good plateaus in the effective energies are obtained for 
relatively high values of the time argument $t$.
For small values of $t$ the eigenvalues show a
qualitatively different behaviour. Therefore
the calculation of generalized eigenvalues is not successful here,
because for the method of generalized eigenvalues the reference time
$t_0$ of equation
(\ref{methods-eq1}) has to be chosen small to reduce 
the statistical errors. 

For the $D_{\pi\pi}(t)$--function the use of smeared sources is not as
advantageous as for the two--pion correlation function $\cpi(t)$.  
Despite smearing the matrix
elements of the subtracted function $D^{LT}_{S,\pi\pi}(t) \cdef
D^{LT}_{\pi\pi}(t) - D^{LT}_{\pi\pi}(t+2)$ for $t=10$ and
$(\beta=10,m_0=0.11)$ have a bad signal--to--noise ratio of about
$0.02$ to $0.08$. The subtracted correlation
function is used here in order to 
avoid any contributions from constant terms.
The signal--to--noise ratio above has to be compared with 
the signal--to--noise ratio of $4.18(2)$ for the
subtracted function $\cspi(t) \cdef \cpi(t) - \cpi(t+2)$ 
with $\alpha = 0.1$. 
Therefore we
had to generate for each point a minimum of $65000$ independent 
configurations to reduce
the statistical errors of the two--pion energies to about $1\%$. 

The comparison of the $D^{LT}_{\pi\pi}(t)$--function with the
correlation function $C^{LT}(t) \equiv \langle O_i^{LT}(t)
O_i^{LT}(0) \rangle$ justifies the usage of the former function: 
The
ratio $D^{LT}_{\pi\pi,ij}(t)/C^{LT}_{ij}(t)$
of the matrix elements of both functions for the parameters
given above is less than $10^{-3}$. 
This means that the $D^{LT}_{\pi\pi}(t)$
parts of the $C^{LT}(t)$--correlation function are strongly
suppressed. Thus it is appropriate to use the correlations  
$D^{LT}_{\pi\pi}(t)$ from the beginning.

\begin{figure}[p]
\vspace*{-1.5cm}
\hspace*{1cm}
\epsfig{file=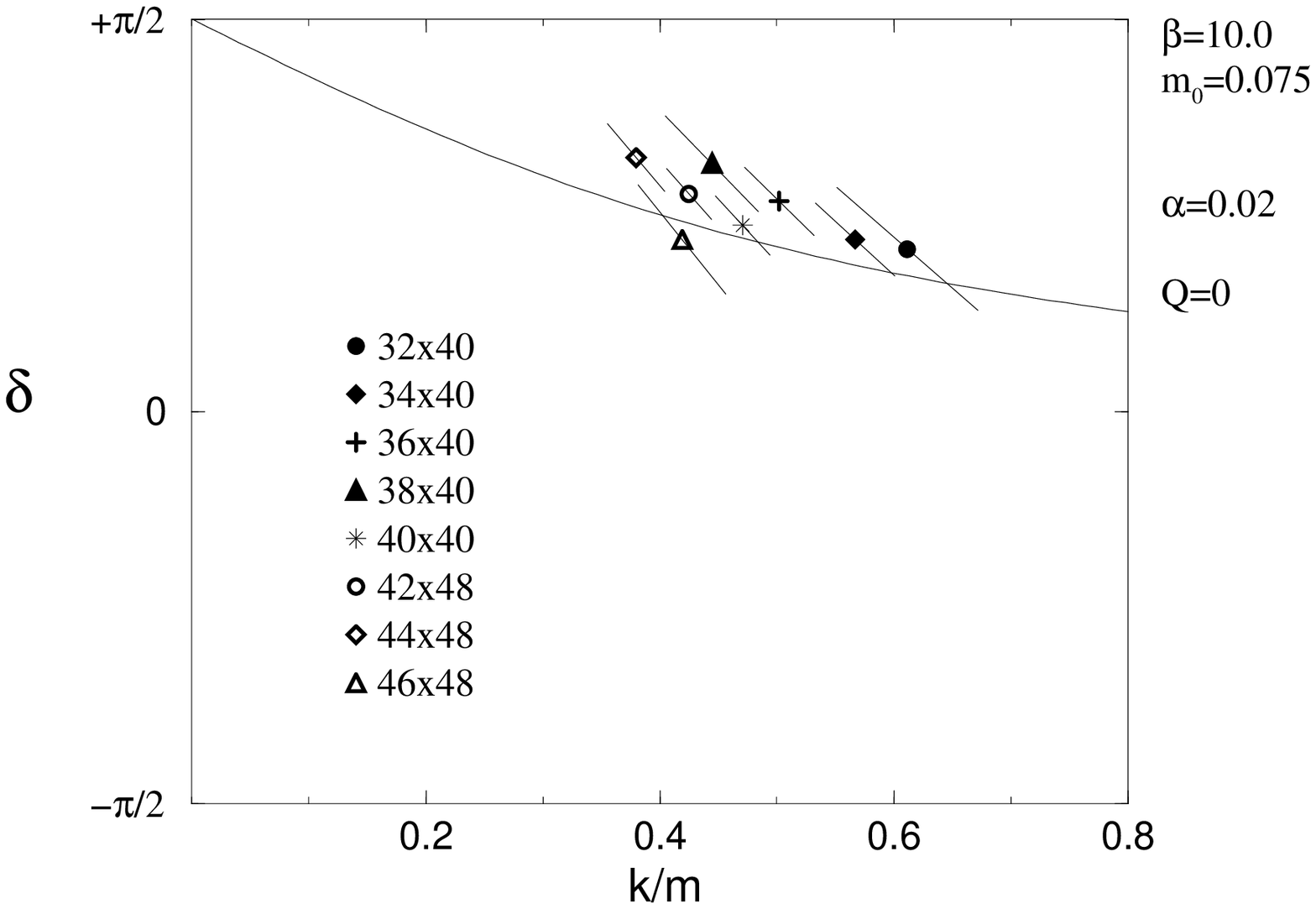,height=0.5\textheight}\\[-2cm]
\vspace*{-1cm}
\hspace*{1cm}
\epsfig{file=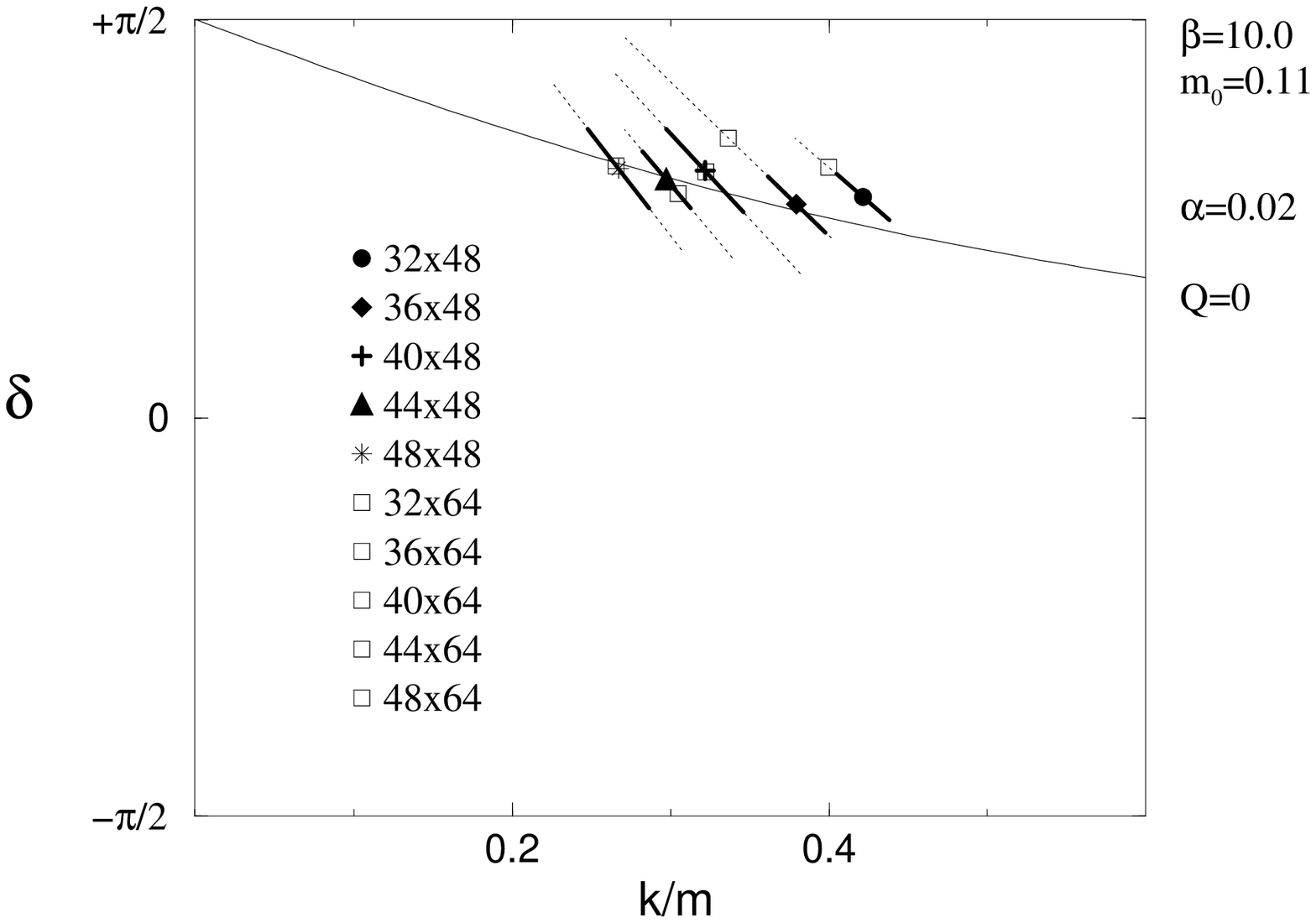,height=0.5\textheight}
\caption{\capsize Scattering phases for two different bare masses and 
different lattice sizes. The analytical
prediction for the $\pi$--$\pi$ scattering according to eq. 
(\ref{schwinger-eq1})
is represented by the solid curve. The mass of
the scattered pions is $m_\pi=0.2545(10)$ and
$m_\pi=0.3367(6)$ respectively. The solid/dotted bars in the lower diagram
are the statistical errors of the data for $T=48/T=64$.}
\label{streunum-fig4}
\end{figure}

\begin{figure}[tb]
\vspace*{-1.5cm}
\hspace*{1cm}
\epsfig{file=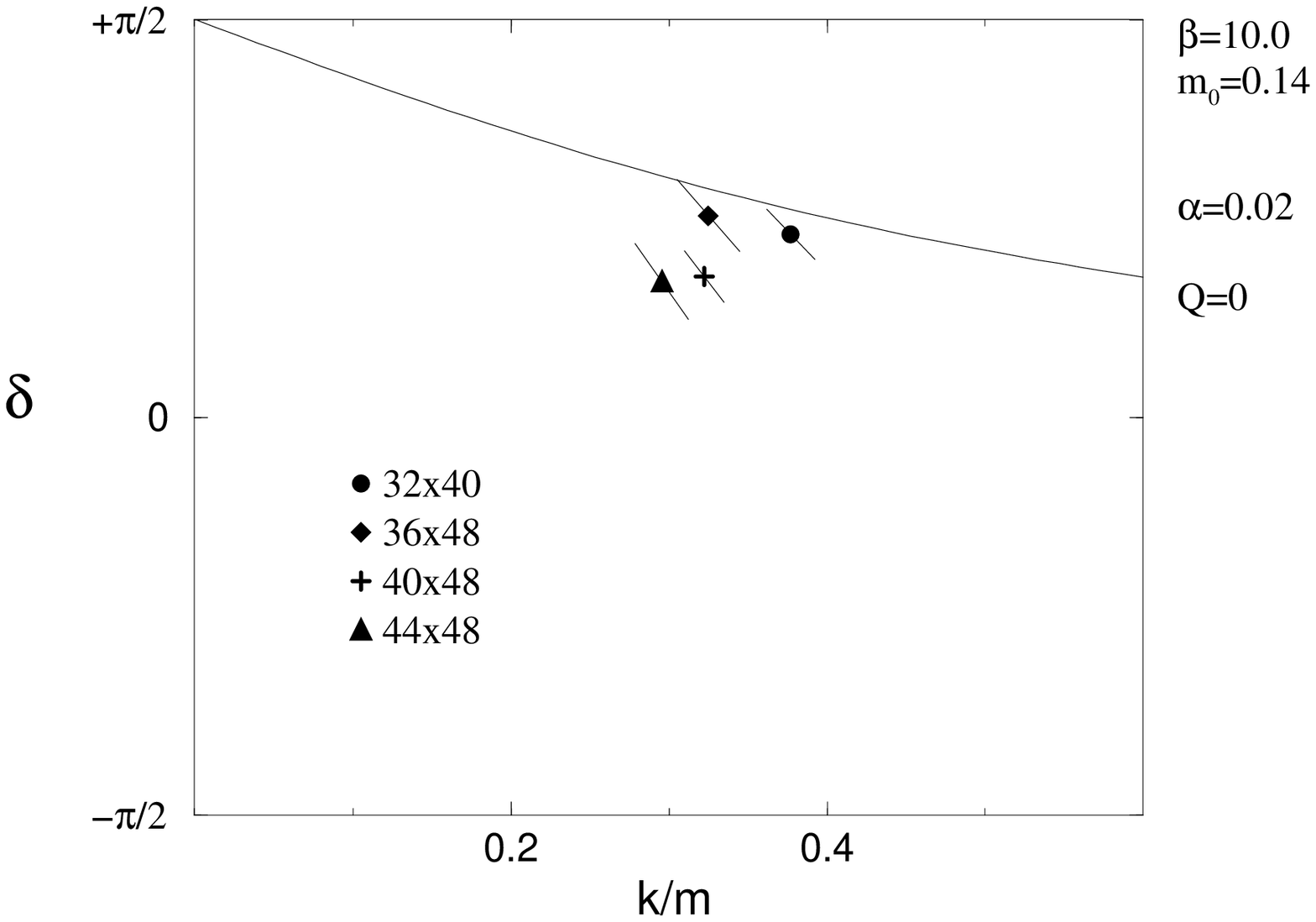,height=0.5\textheight}\\
\capskip
\caption{\capsize Scattering phases for $\beta=10$ and $m_0=0.14$. 
The pion mass is $m_\pi=0.4010(2)$.}
\label{streunum-fig5}
\end{figure}

The scattering phases which were calculated from the
two--particle energies by the bosonic dispersion relation 
(\ref{massnum-eq15}) are shown
in figs. \ref{streunum-fig4} and \ref{streunum-fig5}.
In fig. \ref{streunum-fig4} the numerical values confirm well 
within errors 
the analytical prediction (\ref{schwinger-eq1})
for the $\pi$--$\pi$ scattering. 
This means that in both cases ($\beta=10$, $m_0=0.075$) and 
($\beta=10$, $m_0=0.11$) for 
a ratio $m_0/e \equiv m_0\sqrt{\beta}$ of $0.24$ 
and $0.35$ respectively the strong--coupling prediction describes the
data very well. 
Therefore the analytical strong--coupling prediction for the elastic 
$\pi$--$\pi$ scattering phases in the Schwinger model as resulting
from the sine--Gordon model are confirmed.

If the ratio $m_0/e$ is increased further to
$m_0/e = 0.44$ for ($\beta=10$, $m_0=0.14$) it becomes obvious
from fig. \ref{massnum-fig1}
that the prediction (\ref{schwinger-eq1}) 
is no longer valid.
Indeed the lowest numerical determined scattering phases in fig.
\ref{streunum-fig5} lie
below the analytical curve.  
Considering our numerical results for the mass spectrum
these deviations can {\em not} result from inelastic processes. 
The threshold for
possible inelastic processes like $\pi \pi \rightarrow \eta \eta$ 
or $\pi \pi \rightarrow f_0 f_0$ for the parameters in question is far 
above the determined two--particle energies. Furthermore, we did not
find any particles, e.g. ($I^{PG}=0^{++}$)--mesons, 
which might occur as resonances in the examined energy region and
the lattice sectors considered. Therefore the deviations in
fig. \ref{streunum-fig5} can not be caused by
resonances in the elastic scattering channel.

Because there are no 
inelastic thresholds near or below respectively the calculated energies 
we can conclude that the investigated
pion--pion state is a definite two--particle state which does not
mix with other many--particle states. Therefore according to
section \ref{streunum-luescher} the applicability of
L\"uscher's method is guaranteed. Within the statistical errors the
numerical results do not indicate that it is necessary to consider any
CDD--factors for the elastic pion--pion scattering.
Nevertheless it is possible to multiply 
the ``minimal'' solution (\ref{schwinger-eq1}) 
with CDD--factors of the type
(\ref{schwinger-eq12}), e.g. two factors 
with $\alpha_1 \simeq - \alpha_2$, 
without changing
the shape of the scattering phases qualitatively.

\sect{Summary}

In this paper we present the 
mass spectrum of light mesons in the massive
Schwinger model with an $\tsu_f$ flavour symmetry. Furthermore,
L\"uscher's method is successfully applied to the calculation of
mesonic scattering phases in this model. For strong coupling the
analytical predictions for the pion mass and for the elastic scattering
phases of the pion--pion scattering are confirmed. 

Investigating the mass spectrum in the Schwinger model
it is shown
that for $\beta=10$ and for small bare fermion masses 
$m_0$ the
dependence of the pion mass on the bare parameters corresponds
to the analytical predictions. A significant improvement of the
data is achieved compared to the results for $\beta=4$
\cite{GK98}. This was expected as the continuum limit requires
$\beta \rightarrow \infty$. 

The $f_0$ and $\eta$ singlet states which are analytically predicted 
in addition to the pion are investigated in the strong coupling region, i.e. 
for small values of $m_0 \sqrt{\beta}$. 
The numerical results
for the mass of the $f_0$--meson are for $\beta=5$ and $\beta=7$ 
in the region of the analytically
predicted value of $m_{f_0} = \sqrt{3}\,m_\pi$. 
The mass of the
$\eta$--meson is predicted to be $m_\eta = e \sqrt{2/\pi}$ in the
massless $\tsu_f$ Schwinger model
($e$ is the coupling parameter in the continuum).
We find in the massive case that there are corrections
to this value which are linear in $m_0$.    

Additionally in the mass region
$[m_\pi,4m_\pi]$ we find a rich mass spectrum consisting of 
scalar and pseudoscalar triplet and singlet
states. We assume the lightest of these
particles to be stable. 

The results for the elastic scattering phases of the pion--pion
scattering show that the relevant two--pion states have only a slight
overlap with the operators considered. In fact the signal--to--noise ratio
of the correlation function used is below $10\%$ despite smearing. 
To improve the
separation of the pion--pion energies from other states with large
amplitude we use improved techniques
like specific correlation functions for the two--pion
states and enlarged correlation matrices. In this way it is possible to
obtain scattering phases for the elastic pion--pion scattering using
the method of L\"uscher. For strong coupling these scattering phases are
in good agreement with the analytical predictions.  
 
The experience gained in this project may be useful for the
calculation of elastic scattering phases in other gauge field
theories. It turns out that the investigation of scattering processes
with composite particles like mesonic states
is much more difficult than the scattering of
particles which are no bound states. This is in agreement with investigations
of other fermionic models \cite{FD94,CF97,We97a}. The main problem is
the poor overlap of the two--meson states with the conventional
operators. To compensate for this we had to generate up to $100000$
independent configurations per simulation point. 
Therefore the development of improved operators should be
the main task for investigations of mesonic scattering
processes in QCD by L\"uscher's method, 
e.g. of the $\rho$--resonance in a $\pi$--$\pi$ scattering.

\sect*{Acknowledgement}
We would like to thank M.~G\"ockeler, J.~Jers{\'a}k, C.~B.~Lang
and J.~Westphalen for valuable and helpful discussions. 
We are grateful to J.~Jers{\'a}k 
and his former collaborators
for providing us with an initial version of the HMC program. 
The extensive support
from the NIC J\"ulich and the Rechenzentrum of the RWTH Aachen with
computer time is acknowledged. Finally, we wish to thank
the Deutsche Forschungsgemeinschaft for supporting this project.

\end{document}